\documentclass[11pt]{article}
\usepackage{amsmath,color}
\usepackage{amssymb}
\usepackage{latexsym}
\usepackage{epsfig}

\newcommand{\bigint}{{\scalebox{1.15}{$\displaystyle \int$}}}
\newcommand{\be}{\begin{equation}}
\newcommand{\ee}{\end{equation}}

\newcommand{\bra}[1] {\left<#1\right|}
\newcommand{\ket}[1] {\left|#1\right>}
\newcommand{\braket}[2] {\left<#1\vphantom{#2}\right|\left.\!\vphantom{#1}{#2}\right>}

\newcommand{\cP}{{\cal P}}

\newcommand{\bs}{\begin{subequations}}
\newcommand{\es}{\end{subequations}}
\newcommand{\bbm}{\begin{bmatrix}}
\newcommand{\ebm}{\end{bmatrix}}

\begin{document}

\begin{titlepage}
\rightline{\tt arXiv:1201.1761}
\rightline{\tt MIT-CTP-4332}
\rightline{\tt UT-Komaba/12-1} 
\rightline{\tt TAUP-2935-11}
\begin{center}
\vskip 0.8cm
{\Large \bf {
Open superstring field theory I: gauge fixing, }}\\
\vskip 0.5cm
{\Large \bf {
ghost structure, and propagator
}}\\
\vskip 1.0cm
{\large {Michael Kroyter,${}^1$ Yuji Okawa,${}^2$ 
Martin Schnabl,${}^3$}
\vskip 0.5cm
{\large Shingo Torii${}^2$ and Barton Zwiebach${}^4$}}
\vskip 1cm
${}^1${\it {School of Physics and Astronomy}}\\
{\it {The Raymond and Beverly Sackler Faculty of Exact Sciences}}\\
{\it {Tel Aviv University, Ramat Aviv, 69978, Israel}}\\
\vskip 0.3cm
${}^2${\it {Institute of Physics, University of Tokyo}}\\
{\it {Komaba, Meguro-ku, Tokyo 153-8902, Japan}}\\
\vskip 0.3cm
${}^3${\it {Institute of Physics AS CR}}\\
{\it {Na Slovance 2, Prague 8, Czech Republic}}\\
\vskip 0.3cm
${}^4${\it {Center for Theoretical Physics}}\\
{\it {Massachusetts Institute of Technology}}\\
{\it {Cambridge, MA 02139, USA}}\\
\vskip 0.3cm
mikroyt@tau.ac.il, okawa@hep1.c.u-tokyo.ac.jp, schnabl.martin@gmail.com,\\
storii@hep1.c.u-tokyo.ac.jp, zwiebach@mit.edu
\vskip 0.7cm

\vskip .6cm
{\bf Abstract}
\end{center}

\vskip 0.4cm

\noindent
\begin{narrower}
The WZW form of open superstring field theory has linearized
gauge invariances associated with the BRST operator $Q$
and the zero mode $\eta_0$ of the picture minus-one fermionic 
superconformal ghost.
We discuss gauge fixing of the free theory
in a simple class of gauges
using the Faddeev-Popov method.
We find that the world-sheet ghost number of ghost and antighost string fields ranges
over all integers, except one, and at any fixed ghost number,
only a finite number of picture numbers appear.
We calculate the propagators
in a variety of gauges and determine the field-antifield content
and the free master action in the Batalin-Vilkovisky formalism.
Unlike the case of bosonic string field theory,
the resulting master action is not simply related
to the original gauge-invariant action
by relaxing the constraint on the ghost and picture numbers.

\end{narrower}

\vspace{4cm}

\end{titlepage}

\baselineskip 17pt

\tableofcontents

\section{Introduction and summary}

\newcommand{\p}{\partial}

\newcommand{\hp}{{\Phi}}
\newcommand{\hq}{{Q_B}}
\newcommand{\he}{{\eta_0}}
\newcommand{\ha}{{{A}}}
\newcommand{\rrr}{\big\rangle\big\rangle}
\newcommand{\lllb}{\Bigl\langle\Bigl\langle}
\newcommand{\rrrb}{\Bigr\rangle\Bigr\rangle}

String field theory is an approach to string theory that aims to
address non-perturbative questions that are difficult to study in
the context of first quantization.  Classical solutions that represent 
changes of the open string background are of particular interest,
and  considerable progress was made in this subject in the last few years
(see, for example,~\cite{Fuchs:2008cc,Bonora:2010hi,Kiermaier:2010cf,Erler:2011tc,Noumi:2011kn}).
 
A covariant string field theory should satisfy a series of consistency checks.
The kinetic term, for example, must define the known spectrum of the theory.
The full action, with the inclusion of interaction terms, 
has  nontrivial gauge invariances.
It must be possible to gauge fix these symmetries, derive a propagator,
and set up a perturbation theory that produces off-shell amplitudes that,
on-shell, agree
with the amplitudes in the first-quantized theory.
The purpose of these checks
is not necessarily to construct off-shell amplitudes, but rather to
test the consistency and understand better the structure of the theory.
Indeed that was the way it turned out for open bosonic string
field theory~\cite{Witten:1985cc}.  The Faddeev-Popov quantization of the theory quickly 
suggested that the full set of required ghost and antighost fields could be obtained by relaxing the ghost number constraint on the classical string 
field~\cite{Thorn:1986qj,Bochicchio:1986zj,Bochicchio:1986bd}.\footnote{
In this paper we refer to the string field in the gauge-invariant action before gauge fixing  as the ``classical'' string field,
distinguishing it from ghost and antighost fields introduced by gauge fixing.
}
Moreover, the Batalin-Vilkovisky (BV) quantization approach \cite{Batalin:1981jr,Batalin:1984jr}
turned out to be surprisingly effective~\cite{Thorn:1988hm}.
The full master action for open bosonic 
string field theory---the main object in this quantization scheme---is simply
the classical action evaluated with the unconstrained string field.
For the closed bosonic string field theory, the BV master equation was useful in the construction of the full quantum action, since it has a close
relation with the constraint that ensures proper covering of the moduli spaces of Riemann surfaces~\cite{Zwiebach:1992ie}.
As is the case for open strings, the closed string field theory master action is simply obtained by relaxing the ghost number constraint on the classical string~field.   

It is the purpose of this paper to begin a detailed study of
gauge fixing of the WZW open superstring field theory~\cite{Berkovits:1995ab}
using the Faddeev-Popov method
and the Batalin-Vilkovisky formalism.
This theory describes the Neveu-Schwarz sector of
open superstrings  using the `large' Hilbert space of
the superconformal ghost sector in terms of
$\xi$, $\eta$, and $\phi$~\cite{Friedan:1985ge}.
As opposed to some alternative
formulations~\cite{Witten:1986qs,Preitschopf:fc,Arefeva:1989cp}
no world-sheet insertions of picture-changing operators are required  and the string field theory action  takes the form
\be \label{e0}
S={1\over 2g^2}\Bigl\langle (e^{-\Phi} Q e^{\Phi})
(e^{-\Phi}\eta_0 e^\Phi)
- \int_0^1 dt
(e^{-t\Phi}\p_t e^{t\Phi})\bigl\{ (e^{-t\Phi}Q e^{t\Phi}),
(e^{-t\Phi}\eta_0 e^{t\Phi})\bigr\}\Bigr\rangle\, .
\ee
Here $\{ A, B\} \equiv AB+ BA$,
$g$ is the open string coupling constant,
$\eta_0$ denotes the zero mode of the 
superconformal ghost field $\eta$,
and $Q$ denotes the BRST operator.  
These two operators anticommute and square to zero:
\be \label{etaqprop}
\{ Q, \eta_0\} = 0, \quad Q^2 = \eta_0^2 = 0 \,.
\ee
The string field $\Phi$ is Grassmann
even and has both ghost and picture number zero.  Both $Q$ and $\eta_0$ have
ghost number one.  While $\eta_0$ carries picture number minus one,
$Q$ carries no picture number.
Products of string fields are defined using the star product in~\cite{Witten:1985cc},
and the BPZ inner product of string fields $A$ and $B$ is
denoted by $\langle \, A \, B \, \rangle$ or by $\langle \, A \, | B \, \rangle$.
The action is defined
by expanding all exponentials in formal Taylor series, 
and we evaluate the associated correlators
recalling that  in the large Hilbert space   
\be
\label{basoverlap}
\langle\xi(z) c\p c\p^2 c(w) e^{-2\phi(y)}\rangle \not= 0\, . 
\ee
The  action can be shown to be invariant under  gauge
transformations with infinitesimal gauge parameters $\Lambda$ and $\Omega$:
\be \label{egtrs}
\delta e^{\Phi} = (Q \Lambda) e^{\Phi} + e^{\Phi}(\eta_0\Omega) \, ,
\ee
and the equation of motion for the string field is
\be\label{eom}
\eta_0(e^{-\Phi}Q e^{\Phi}) =0 \,.
\ee

In this paper we focus on the linearized theory.
For notational
simplicity we will simply set the open string coupling equal to one: $g=1$.
To linearized
order the action
reduces to $S_0$ given by
\be \label{kinterm}
S_0   \ =\   {1\over 2} \bigl\langle
    \,(Q \hp)\,(\eta_0 \hp)\, \bigr\rangle
     \,.
\ee
Using bra and ket notation, the kinetic term can be written as
\be
S_0 \ = \  - \frac{1}{2}\bra{ \Phi_{(0,0)} } Q \eta_0 \ket{\Phi_{(0,0)}}  \,.
\ee
Here 
we have written  $\Phi= \Phi_{(0,0)}$ to emphasize that the
classical string field has both ghost number and picture number zero.
Unless indicated otherwise, 
we take $X_{(g,p)}$ to be an object that carries ghost number
$g$ and picture number $p$.
To this order the equation of motion (\ref{eom}) becomes
\begin{equation}
\label{linfeqn}
\eta_0 Q \,\Phi_{(0,0)}\ = \ 0 \,, 
\end{equation}
and the gauge transformations (\ref{egtrs}) become
\be
\label{linearized-gauge-transformations}
\delta_0 \Phi_{(0,0)} = Q \Lambda \,+\,  \eta_0\Omega\,.
\ee
Let us use $\epsilon$ for gauge parameters and rewrite (\ref{linearized-gauge-transformations}) as
\be
\label{gtzero}
\delta_0 \Phi_{(0,0)} = Q \epsilon_{(-1,0)} \,+\,  \eta_0\epsilon_{(-1, 1)}\,,
\ee
where we have indicated the appropriate ghost and picture numbers in
the subscripts.  Note that both $\epsilon_{(-1,0)}$ and 
$\epsilon_{(-1, 1)}$ are Grassmann odd, both have ghost number minus one, but
differ in picture number. The gauge invariances (\ref{gtzero}) have their own gauge invariances.
We can change $\epsilon_{(-1,0)}$ and $\epsilon_{(-1,1)}$ without 
changing $\delta_0 \Phi_{(0,0)}$.  Indeed, with 
\be
\begin{split}
\delta_1 \epsilon_{(-1,0)}  &=  Q \epsilon_{(-2,0)}  + \eta_0 \epsilon_{(-2,1) } \,, \\[0.5ex] 
\delta_1 \epsilon_{(-1,1)}&=  Q \epsilon_{(-2,1)}  + \eta_0 \epsilon_{(-2,2) }\,,
\end{split}
\ee
we readily verify that $\delta_1 (\delta_0  \Phi_{(0,0)} )= 0$, making 
use of~(\ref{etaqprop}).  At this stage we have introduced three gauge
parameters, all of ghost number minus two, and with pictures zero, one, and
two.   The above redundant transformations have
their own redundancy:
\be
\begin{split}
\delta_2 \epsilon_{(-2,0)}  &=  Q \epsilon_{(-3,0)}  + \eta_0 \epsilon_{(-3,1) } \,, \\[0.5ex] 
\delta_2 \epsilon_{(-2,1)}&=  Q \epsilon_{(-3,1)}  + \eta_0 \epsilon_{(-3,2) } \,, \\[0.5ex] 
\delta_2 \epsilon_{(-2,2)}&=  Q \epsilon_{(-3,2)}  + \eta_0 \epsilon_{(-3,3) }\,,
\end{split}
\ee
and this time we verify that 
$\delta_2 (\delta_1  \epsilon_{(-1,0)} )\, = \,
\delta_2 (\delta_1  \epsilon_{(-1,1)} )= 0 $. 
At step $n$, in matrix notation, we have
\begin{equation}
\label{infredgauge}
\delta_n \left(
\begin{array}{c}
\epsilon_{(-n,0)} \\
\epsilon_{(-n,1)} \\
\epsilon_{(-n,2)} \\
\vdots \\
\epsilon_{(-n,n)}
\end{array}
\right) \ = \ \left(
\begin{array}{cccccc}
Q & \eta_0 & 0 & \cdots & 0 & 0 \\[0.1ex]
0 & Q & \eta_0 & \cdots & 0 & 0 \\[0.1ex]
0 & 0 & Q & \cdots & 0 & 0 \\[0.1ex]
\vdots & \vdots & \vdots & \ddots & \vdots & \vdots \\
0 & 0 & 0 & \cdots & Q & \eta_0
\end{array}
\right)
\left(
\begin{array}{c}
\epsilon_{(-(n+1),0)} \\
\epsilon_{(-(n+1),1)} \\
\epsilon_{(-(n+1),2)} \\
\vdots \\
\epsilon_{(-(n+1),n)}\\
\epsilon_{(-(n+1),n+1)}
\end{array}
\right) \,. 
\end{equation}
The above describes the full structure of redundant symmetries of the theory at linearized level.   It is the starting point for the BRST quantization of the theory, where we select gauge-fixing conditions and add suitable
Faddeev-Popov terms to the action.  The above gauge parameters turn into
ghosts 
\be
\Phi_{(-n, p)}  \,,  ~~  n\geq 1\,, ~~ p = 0,1, \ldots , n\,.
\ee
It follows from the BRST prescription that all of the above ghost
fields are Grassmann even, just like the classical string field
$\Phi_{(0,0)}.$\footnote{
The spacetime fields in such string 
fields can be even or odd depending
on the Grassmann parity
of the CFT basis states.} Antighosts must be also added.    The Faddeev-Popov quantization
is carried out using a set of gauge conditions that enable us to confirm
that the free gauge-fixed action, after elimination of auxiliary fields,
coincides with that of Witten's free theory~\cite{Witten:1986qs} in Siegel gauge.
The gauge-fixing conditions
here are of type $(b_0, \xi_0; \alpha)$, meaning that 
ghosts and antighosts are required to be annihilated
by operators made of the zero modes $b_0$ and $\xi_0$
with a parameter $\alpha$. (See~(\ref{one-param bxi}).)

We then turn to the calculation of the propagator of the theory, for which
the free gauge-fixed action
is sufficient. 
As usual, we add to the action linear couplings that associate unconstrained sources 
with the classical field, 
with each ghost, and 
with each antighost.  The propagator is then the matrix that defines the
quadratic couplings of sources in the action, and it is obtained by solving
for all fields in terms of sources using the classical equations of motion.  We examine this propagator
for a few types of gauges.  In the $(b_0, \xi_0; \alpha)$ type gauges, the
propagator matrix contains
the zero mode $X_0 = \{ Q , \xi_0\}$
of the picture-changing operator
and its powers. 
The propagator
is quite complicated for $\alpha \not= 0$ and simplifies somewhat for
$\alpha = 0$, where it takes the form of matrices of triangular type.  

A more intriguing class of gauges are of type $(b_0, d_0;\alpha)$.  
Here $d_0$ is the zero mode of the operator 
$d = [ Q , b\xi ]$.
In the language of the twisted $N=2$ superconformal 
algebra~\cite{Berkovits-Vafa}, 
$d_0 = \widetilde G^-_0$
is a counterpart of
$b_0 = G^-_0$.
Corresponding to the relation $\{Q, b_0\} = L_0$, 
the anticommutation relation $\{ \eta_0, d_0\} = L_0$ holds.
In fact, the gauge-fixing conditions $b_0 \Phi_{(0,0)} = d_0 \Phi_{(0,0)} = 0$ were used
in the calculation of a four-point amplitude in~\cite{Berkovits:1999bs}.
The propagators in this class of gauges are much simpler
than in the $(b_0, \xi_0;\alpha)$ type gauges
and do not involve picture-changing operators.
They further simplify when   $\alpha = 1$ (see 
(\ref{doxioact}), (\ref{doxioSn}), and (\ref{doxioPk})).  
We expect this form of the propagator to be useful 
in the study of loop amplitudes. 

The gauge structure of the free theory is infinitely reducible.  In fact,
the equations in (\ref{infredgauge}) determine the ``field/antifield" structure
of the theory following the usual Batalin-Vilkovisky procedure
\cite{Batalin:1981jr,Batalin:1984jr} (reviewed in
\cite{Henneaux:1989jq,Henneaux:1992ig,Gomis:1994he}).  We write  the
original gauge symmetry of the classical fields $\phi^{\alpha_0}$ schematically as 
\be
\begin{split}
\delta \phi^{\alpha_0}  \ = \  R_{(0)\alpha_1}^{\,\alpha_0}  \,  \epsilon^{\alpha_1} \,,
\end{split}
\ee
where sum over repeated indices is implicit and  $R$ is  possibly field dependent. The symmetry is infinitely reducible if there are 
gauge invariances of gauge invariances at every stage, namely
\be
\begin{split}
\delta \epsilon^{\alpha_1} &  \ = \  R_{(1)\alpha_2}^{\,\alpha_1}  \,  \epsilon^{\alpha_2} \\
\delta \epsilon^{\alpha_2} &  \ = \  R_{(2)\alpha_3}^{\,\alpha_2}  \,  \epsilon^{\alpha_3} \\
\vdots  ~~& \ = \  \quad \vdots 
\end{split}
\ee
with the following on-shell relations
\be
R_{(n)\alpha_{n+1}}^{\,\alpha_{n}} \, R_{(n+1)\alpha_{n+2}}^{\,\alpha_{n+1}} \ = \ 0 \,,
~~\hbox{for}~~n = 0 , 1,  2, \ldots \,. 
\ee
In this case one introduces fields $\phi^{\alpha_n}$
with $n\geq 1$
and antifields $\phi^*_{\alpha_n}$ with $n\geq 0$ such that the
BV action reads
\be
\label{boundcond}
\begin{split}
S \ &= \   S_0(\phi^{\alpha_0})  +  \sum_{n=0}^\infty
 \phi^*_{\alpha_n} R_{(n)\alpha_{n+1}}^{\,\alpha_n}  \phi^{\alpha_{n+1}}   +\ldots \\
 \ &= \  S_0 + \phi^*_{\alpha_0} R_{(0)\alpha_1}^{\,\alpha_0}\phi^{\alpha_1}
 + \phi^*_{\alpha_1} R_{(1)\alpha_2}^{\,\alpha_1} \phi^{\alpha_2}  + \ldots 
 ~,
\end{split}
\ee
where the dots 
represent terms at least cubic in ghosts and antifields  that are
needed for a complete solution of the master equation.
Since all string fields for ``fields"
in open superstring field theory
are Grassmann even,
the $R$'s are Grassmann odd, and since 
the inner product with (\ref{basoverlap})
needed
to form the action couples states of the same
Grassmann parity, the string fields for ``antifields"
are Grassmann odd.\footnote{In open bosonic  
string field theory the string fields for fields and 
those for
antifields are of the same (odd) Grassmann parity.}
The antifield $\Phi_{(g,p)}^*$associated with the
field $\Phi_{(g,p)}$ is $\Phi_{(2-g,-1-p)}$:
\be
\label{bvbrack}
 \Phi_{(g,p)}^*  \ = \  \Phi_{(2-g, -1-p)} \,.
\ee
This follows from 
(\ref{boundcond}) where each term in the sum takes the form 
$\phi^*_{\alpha_n} (\delta \phi^{\alpha_n})$, with the gauge parameter
replaced by a ghost field of the same ghost and picture number.  
This implies that
the inner product with (\ref{basoverlap})
must be able to
couple a field to its antifield. Since this inner product
requires a total ghost number violation of two and a total picture
number violation of minus one, the claim in (\ref{bvbrack}) follows.    
The full field/antifield structure of the theory is therefore
\begin{equation}
\begin{split}\\
& \cdots ~~\Phi_{(-2,2)} \\
&\cdots~~  \Phi_{(-2,1)} ~~ \Phi_{(-1,1)} \hskip180pt \Big\uparrow p \\
&\cdots ~~ \Phi_{(-2,0)} ~~ \Phi_{(-1,0)} ~~ \Phi_{(0,0)}  ~~ ~-~ ~~ 
\hskip120pt  \longrightarrow g \\
&\phantom{\cdots ~~ \Phi_{(-2,0)} ~~ \Phi_{(-1,0)} ~~ \Phi_{(0,0)}} \hskip20pt ~~~~~~~ \Phi_{(2,-1)}~~\Phi_{(3,-1)} ~~\Phi_{(4,-1)} ~~\cdots \\
& \phantom{\cdots ~~\phantom{ \Phi_{(-2,0)} ~~ \Phi_{(-1,0)} ~~ \Phi_{(0,0)}} \hskip20pt ~~~~~~~ \Phi_{(2,-1)}} ~~\Phi_{(3,-2)} ~~\Phi_{(4,-2)} ~~\cdots\\
&\phantom{\cdots~~ \phantom{\phantom{ \Phi_{(-2,0)} ~~ \Phi_{(-1,0)} ~~ \Phi_{(0,0)}} \hskip20pt ~~~~~~~ \Phi_{(2,-1)}} ~~\Phi_{(3,-2)}}~~\Phi_{(4,-3)} ~~ \cdots\\[0.5ex]\\
\end{split}
\end{equation}
The string fields $\Phi_{(g,p)}$ with $g\leq 0$ on the left side
are the ``fields," 
and the $\Phi_{(g,p)}$ with $g\geq 2$ on the right side
are the ``antifields."  
Note the gap at
$g=1$.   Collecting  
all the fields in $\Phi_-$ 
and all antifields in $\Phi_+$ as 
\be
\Phi_-\ = \ \sum_{g = 0}^\infty \sum_{p= 0}^g   \Phi_{(-g,p)}\,,
~~~~\Phi_+\ = \ \sum_{g=2}^\infty\sum_{p=1}^{g-1}  \Phi_{(g,-p)}\,,
\label{Phi+-} 
\ee
we can show that the
free master action $S$ 
implied by (\ref{boundcond}) and by our identification of fields
and antifields 
takes the form:
\begin{equation}
\label{free-master}
S \ =  \ - \, \frac{1}{2}\bra{\Phi_-} Q \eta_0 \ket{\Phi_-}
  + \bra{\Phi_+} (Q + \eta_0) \ket{\Phi_-}\,.
\end{equation}
The master equation $\{ S, S\} =  0$, where $\{ \cdot \,, \, \cdot\}$
is the BV antibracket, will be shown to be satisfied.

\section{Gauge fixing of the free theory}
\setcounter{equation}{0}

In this section
we perform gauge fixing 
of the free open superstring field theory
using the Faddeev-Popov method. 
We first review the procedure in the free open bosonic  string field theory,
and then we extend it to open superstring field theory.
We also demonstrate that the resulting gauge-fixed action
coincides with that of Witten's superstring field theory in Siegel gauge
after integrating out auxiliary fields.

\subsection{Open bosonic string field theory}

The gauge-invariant action of the free theory is given by
\begin{equation}
S_0 = -\frac{1}{2} \langle \Psi_1 | Q | \Psi_1 \rangle \,,
\end{equation}
where $\Psi_1$ is the open string field.
It is Grassmann odd and carries ghost number one,
as indicated by the subscript.
The BRST operator $Q$ is BPZ odd: $Q^\star = -Q$.
This action is invariant under the following gauge transformation:
\begin{equation}
\delta_\epsilon \Psi_1 = Q \epsilon_0 \,,
\end{equation}
where $\epsilon_0$ is a Grassmann-even string field of ghost number zero.

The Faddeev-Popov method
consists of adding two terms
to the gauge-invariant action.
The first term is given by
\begin{equation}
\mathcal{L}_{\rm GF} = \lambda^i F_i (\phi) \,,
\end{equation}
where $F_i (\phi)=0$ are the gauge-fixing conditions on the field $\phi$
and $\lambda^i$ are the corresponding Lagrange multiplier fields.
The second term is the Faddeev-Popov term given by
\begin{equation}
\mathcal{L}_{\rm FP}
= b^i \, \Bigl( \, c^\alpha \frac{\delta}{\delta \epsilon^\alpha} \, \Bigr) \,
\delta_\epsilon F_i (\phi) \,.
\end{equation}
It is obtained from $\mathcal{L}_{\rm GF}$
by changing $\lambda^i$ to the antighost fields $b^i$
and by changing $F_i (\phi)$ to its gauge transformation $\delta_\epsilon F_i (\phi)$
with the gauge parameters
$\epsilon^\alpha$ replaced by the ghost fields $c^\alpha$.
The sum of the two terms $\mathcal{L}_{\rm GF}+\mathcal{L}_{\rm FP}$
is then BRST exact:
$\mathcal{L}_{\rm GF}+\mathcal{L}_{\rm FP}
= -\delta_B ( \, b^i F_i (\phi) \, )$
under the convention $\delta_B b^i = -\lambda^i$.

Let us apply this procedure to the free theory of open bosonic string field theory
and choose the Siegel gauge condition
\begin{equation}
b_0 \Psi_1 = 0
\end{equation}
for gauge fixing.
Note that $b_0$ is BPZ even: $b_0^\star = b_0$.
It is convenient to decompose $\Psi_1$
into two subsectors according to the zero modes $b_0$ and $c_0$ as follows:
\begin{equation}
\Psi_1 =  \Psi_1^- +  c_0 \Psi_1^c \,,
\end{equation}
where $\Psi_1^-$ and  $\Psi_1^c$ are both annihilated by $b_0$.
The superscript `$-$' indicates the sector without $c_0$
and the superscript `$c$' 
indicates the sector with $c_0$,
although $c_0$ has been removed in $\Psi_1^c$.
Therefore $\Psi_1^c$ is Grassmann even and carries ghost number zero,
and so the subscript, which is carried over from $\Psi_1$,
does not coincide with the ghost number of $\Psi_1^c$.
The operator $c_0$ we used in the decomposition is BPZ odd: $c_0^\star = -c_0$.
Using this decomposition, the Siegel gauge condition can be stated as
\begin{equation}
\Psi_1^c = 0 \,.
\end{equation}
The gauge-fixing term $S_{\rm GF}$ implementing this condition
can be written as
\begin{equation}
S_{\rm GF} = -\langle N | c_0 | \Psi_1^c \rangle \,,
\end{equation}
where the Lagrange multiplier field $N$ is annihilated by $b_0$.
Note that the insertion of $c_0$ is necessary for the inner product
to be nonvanishing.
The ghost number of $N$ is two
and component fields playing the role of Lagrange multiplier fields
have to be Grassmann even, so the string field $N$ is Grassmann even.
This term can be equivalently written as
\begin{equation}
S_{\rm GF} = -\langle N_2 | \Psi_1 \rangle  \,,
\end{equation}
with the constraint
\begin{equation}
b_0 N_2 = 0 \,.
\end{equation}
This can be seen by decomposing $N_2$
before imposing the constraint 
as
\begin{equation}
N_2 =  N_2^- +  c_0 N_2^c \,,
\end{equation}
where $N_2^-$ and $N_2^c$ are annihilated by $b_0$.
The inner product $\langle N_2 | \Psi_1 \rangle$ is then given by
\begin{equation}
\langle N_2 | \Psi_1 \rangle
= \langle N_2^- | c_0 | \Psi_1^c \rangle + \langle N_2^c | c_0 | \Psi_1^- \rangle \,.
\end{equation}
The constraint $b_0 N_2 = 0$ eliminates $N_2^c$,
and the remaining field $N_2^-$ is identified with the Lagrange multiplier field $N$.
The string field $N_2$ is Grassmann even and carries ghost number two.

Another way to derive $S_{\rm GF}$ is to use the form $b_0 \Psi_1 = 0$
for the gauge-fixing condition and write
\begin{equation}
S_{\rm GF} = \langle \widetilde{N}_3 | b_0 | \Psi_1 \rangle \,.
\end{equation}
We then redefine the Lagrange multiplier as
\begin{equation}
N_2 = b_0 \widetilde{N}_3 \,.
\end{equation}
The resulting field $N_2$ is subject to the constraint $b_0 N_2 = 0$.
Since $\{ b_0, c_0 \} = 1$, any solution $N_2$ to this constraint
can be written as $N_2 = \{ b_0, c_0 \} N_2 = b_0 c_0 N_2 = b_0 \widetilde{N}_3$
with $\widetilde{N}_3 = c_0 N_2$.
Therefore, $N_2$ obtained from the redefinition $N_2 = b_0 \widetilde{N}_3$
is equivalent to $N_2$ with the constraint $b_0 N_2 = 0$.

The Faddeev-Popov term $S_{\rm FP}$ can be obtained from $S_{\rm GF}$
by changing $N_2$ to the Grassmann-odd antighost field $\Psi_2$ of ghost number two
and by changing $\Psi_1$ to its gauge transformation
$Q \epsilon_0$ with $\epsilon_0$
replaced by the Grassmann-odd ghost field $\Psi_0$ of ghost number zero.
We have
\begin{equation}
S_{\rm FP} = -\langle \Psi_2 | Q | \Psi_0 \rangle  \,,
\end{equation}
with the constraint
\begin{equation}
b_0 \Psi_2 = 0 \,,
\end{equation}
which is inherited from $b_0 N_2 = 0$.
After integrating out $N_2$, the total action we obtain is
\begin{equation}
S_0 + S_1 = -\frac{1}{2} \langle \Psi_1 | Q | \Psi_1 \rangle
-\langle \Psi_2 | Q | \Psi_0 \rangle \,,
\end{equation}
with
\begin{equation}
b_0 \Psi_1 = 0 \,, \qquad b_0 \Psi_2 = 0 \,.
\end{equation}

This action $S_0+S_1$ is invariant under the following gauge transformation:
\begin{equation}
\delta_\epsilon \Psi_0 = Q \epsilon_{-1} \,.
\end{equation}
We can choose
\begin{equation}
b_0 \Psi_0 = 0
\end{equation}
for gauge fixing.
Repeating the same Faddeev-Popov procedure, 
we obtain
\begin{equation}
S_0 + S_1 + S_2
= -\frac{1}{2} \langle \Psi_1 | Q | \Psi_1 \rangle
-\langle \Psi_2 | Q | \Psi_0 \rangle
-\langle \Psi_3 | Q | \Psi_{-1} \rangle\,,
\end{equation}
with
 \begin{equation}
 b_0 \Psi_1 = 0 \,, \quad b_0 \Psi_2 = 0 \,, \qquad
 b_0 \Psi_0 = 0 \,, \quad b_0 \Psi_3 = 0 \,,
 \end{equation}
where $\Psi_3$ has ghost number three 
and $\Psi_{-1}$ has ghost number minus one. 

The action $S_0+S_1+S_2$ is invariant under
$\delta_\epsilon \Psi_{-1} = Q \epsilon_{-2}$.
In this way the gauge-fixing procedure continues, and at the end we obtain
\begin{equation}
S = \sum_{n=0}^\infty S_n \,,
\end{equation}
where
\begin{equation}
S_0 = -\frac{1}{2} \langle \Psi_1 | Q | \Psi_1 \rangle \,, \qquad
S_n = -\langle \Psi_{n+1} | Q | \Psi_{-n+1} \rangle \quad \text{for} \quad n \ge 1
\end{equation}
with
\begin{equation}
b_0 \Psi_n = 0 \,, ~~~\forall n \,.
\end{equation}
The action $S$ can also be written compactly as
\begin{equation}
S = -\frac{1}{2} \langle \Psi | Q | \Psi \rangle
\quad \text{with} \quad
\Psi = \sum_{n=-\infty}^\infty \Psi_n \,, \quad b_0 \Psi = 0 \,.
\end{equation}

\subsection{Open superstring field theory}
\label{open ssft}
Let us now
perform gauge fixing of the free open superstring field theory.
We denote a string field of ghost number $g$ and picture number $p$
by $\Phi_{(g,p)}$.
The gauge-invariant action of the free theory is given by
\begin{equation}
\label{s0given}
S_0 = -\frac{1}{2} \langle \Phi_{(0,0)} | Q \eta_0 | \Phi_{(0,0)} \rangle \,,
\end{equation}
where $\eta_0$ is the zero mode of the superconformal ghost
carrying ghost number one 
and picture number minus one. 
It is therefore BPZ odd: $\eta_0^\star = -\eta_0$.
The action $S_0$ is invariant under the following gauge transformations:
\begin{equation}
\delta_{\epsilon} \Phi_{(0,0)} = Q \epsilon_{(-1,0)} + \eta_0 \epsilon_{(-1,1)} \,.
\end{equation}
We can choose $\epsilon_{(-1,1)}$ appropriately
such that the condition
\begin{equation}
\xi_0 \Phi_{(0,0)} = 0
\label{xi_0-Phi_(0,0)} 
\end{equation}
on $\Phi_{(0,0)}$ is satisfied.
The operator $\xi_0$ we used in the gauge-fixing condition
is BPZ even: $\xi_0^\star = \xi_0$.
Because $\{ \eta_0, \xi_0 \} =1$,
a field  $\Phi_{(0,0)}$ satisfying  
(\ref{xi_0-Phi_(0,0)})
can be written as
\begin{equation}
\Phi_{(0,0)} = \xi_0 \widehat{\Phi}_{(1,-1)} \,,
\end{equation}
where $\widehat{\Phi}_{(1,-1)}$ carrying ghost number one 
and picture number minus one 
is in the small Hilbert space, namely, it is annihilated by $\eta_0$.
The equation of motion
\begin{equation}
Q \eta_0 \Phi_{(0,0)} = 0
\end{equation}
in the large Hilbert space
reduces to
\begin{equation}
Q \widehat{\Phi}_{(1,-1)} = 0 \,.
\end{equation}
Since this is the familiar equation of motion in the small Hilbert space,
we know that we can choose the condition
\begin{equation}
b_0 \widehat{\Phi}_{(1,-1)} = 0
\end{equation}
to fix the remaining gauge symmetry. 
This gauge-fixing condition can be stated for the original field $\Phi_{(0,0)}$
as $b_0 \Phi_{(0,0)} = 0$ when $\xi_0 \Phi_{(0,0)} = 0$ is imposed.
The condition $b_0 \Phi_{(0,0)} = 0$
can be satisfied by appropriately choosing $\epsilon_{(-1,0)}$,
and it is compatible with $\xi_0 \Phi_{(0,0)} = 0$ by adjusting $\epsilon_{(-1,1)}$.
To summarize, we can choose
\begin{equation}
b_0 \Phi_{(0,0)} = 0 \,, \quad \xi_0 \Phi_{(0,0)} = 0
\label{conditions-on-Phi0}
\end{equation}
as the gauge-fixing conditions on $\Phi_{(0,0)}$.

It is convenient to decompose $\Phi_{(g,p)}$ 
into four subsectors
according to the zero modes $b_0$, $c_0$, $\eta_0$, and $\xi_0$ 
as follows:
\begin{equation}
\Phi_{(g,p)} = \Phi_{(g,p)}^{--} +  c_0 \Phi_{(g,p)}^{c-}
+\xi_0 \Phi_{(g,p)}^{-\xi} +  c_0 \xi_0 \Phi_{(g,p)}^{c \xi} \,,
\label{decomposition}  
\end{equation}
where
$\Phi_{(g,p)}^{--}$, $\Phi_{(g,p)}^{c-}$, $\Phi_{(g,p)}^{-\xi}$, and  $\Phi_{(g,p)}^{c \xi}$
are all annihilated by
both $b_0$ and $\eta_0$.
Note that the subscript $(g,p)$ is carried over from $\Phi_{(g,p)}$
and does not indicate the ghost  and picture numbers of the fields
in the subsectors.
The ghost and picture numbers $(g,p)$ are
$(g,p)$ for $\Phi_{(g,p)}^{--}$,
$(g-1,p)$ for $\Phi_{(g,p)}^{c-}$,
$(g+1,p-1)$ for $\Phi_{(g,p)}^{-\xi}$,
and $(g,p-1)$ for $\Phi_{(g,p)}^{c \xi}$.
Using this notation, the gauge-fixing conditions~(\ref{conditions-on-Phi0}) can be stated as
\begin{equation}
\Phi_{(0,0)}^{--} = 0 \,, \quad \Phi_{(0,0)}^{c-} = 0 \,, \quad \Phi_{(0,0)}^{c \xi} = 0 \,.
\end{equation}
The gauge-fixing term $S_{\rm GF}$ implementing these conditions can be written as
\begin{equation}
S_{\rm GF} =
{}-\langle N_{(2,-1)}^{c \xi} | c_0 \xi_0 | \Phi_{(0,0)}^{--} \rangle
+\langle N_{(2,-1)}^{-\xi} | c_0 \xi_0 | \Phi_{(0,0)}^{c-} \rangle
+\langle N_{(2,-1)}^{--} | c_0 \xi_0 | \Phi_{(0,0)}^{c \xi} \rangle \,,
\end{equation}
where the Lagrange multiplier fields $N_{(2,-1)}^{c \xi}$, $N_{(2,-1)}^{-\xi}$, and $N_{(2,-1)}^{--}$
are all annihilated by $b_0$ and $\eta_0$.
Note that the insertion of $c_0 \xi_0$ to each term is necessary
for the inner product to be nonvanishing,
as can be seen from~(\ref{basoverlap}). 
This term can be equivalently written as
\begin{equation}
S_{\rm GF} = \langle N_{(2,-1)} | \Phi_{(0,0)} \rangle
\end{equation}
with the constraint
\begin{equation}
b_0 \xi_0 N_{(2,-1)} = 0 \,.
\end{equation}
This can be seen by writing
$N_{(2,-1)}$ before imposing the constraint as 
\begin{equation}
N_{(2,-1)} = N_{(2,-1)}^{--} +  c_0 N_{(2,-1)}^{c-}
+\xi_0 N_{(2,-1)}^{-\xi} +  c_0 \xi_0 N_{(2,-1)}^{c \xi} \,,
\label{N_(2,-1)-decomposition}
\end{equation}
where $N_{(2,-1)}^{--}$, $N_{(2,-1)}^{c-}$, $N_{(2,-1)}^{-\xi}$,
and $N_{(2,-1)}^{c \xi}$ are all annihilated by both $b_0$ and $\eta_0$.
The constraint $b_0 \xi_0 N_{(2,-1)} = 0$ eliminates $N_{(2,-1)}^{c-}$,
and the remaining fields
$N_{(2,-1)}^{c \xi}$, $N_{(2,-1)}^{-\xi}$, and $N_{(2,-1)}^{--}$
implement
$\Phi_{(0,0)}^{--} = 0$, $\Phi_{(0,0)}^{c-} = 0$,  and $\Phi_{(0,0)}^{c \xi} = 0$.

Another way to derive $S_{\rm GF}$ is to use the form $b_0 \Phi_{(0,0)} = \xi_0 \Phi_{(0,0)} = 0$
for the gauge-fixing conditions and write
\begin{equation}
S_{\rm GF} =
{}-\langle \widetilde{N}_{(3,-1)} | b_0 | \Phi_{(0,0)} \rangle
-\langle \widetilde{N}_{(3,-2)} | \xi_0 | \Phi_{(0,0)} \rangle \,.
\end{equation}
We then redefine the Lagrange multiplier as
\begin{equation}
N_{(2,-1)} = b_0 \widetilde{N}_{(3,-1)}
+\xi_0 \widetilde{N}_{(3,-2)} \,.
\end{equation}
The resulting field $N_{(2,-1)}$ is subject to the constraint $b_0 \xi_0 N_{(2,-1)} = 0$,
and the solution to the constraint can be written as
$N_{(2,-1)} = b_0 \widetilde{N}_{(3,-1)} +\xi_0 \widetilde{N}_{(3,-2)}$.
This time, however, $\widetilde{N}_{(3,-1)}$ and $\widetilde{N}_{(3,-2)}$
are not uniquely determined for a given solution. 
Comparing this with the decomposition~(\ref{N_(2,-1)-decomposition}),
we find that $N_{(2,-1)}^{--}$ is in the part $b_0 \widetilde{N}_{(3,-1)}$
and $c_0 \xi_0 N_{(2,-1)}^{c \xi}$ is in the part $\xi_0 \widetilde{N}_{(3,-2)}$,
but $\xi_0 N_{(2,-1)}^{-\xi}$ 
can be in either part.
This ambiguity is related to the fact that a part of $\widetilde{N}_{(3,-1)}$
and a part of $\widetilde{N}_{(3,-2)}$ impose the same constraint
$\Phi_{(0,0)}^{c-} = 0$.
More specifically, if we write $\widetilde{N}_{(3,-1)}$ and $\widetilde{N}_{(3,-2)}$ as
\begin{equation}
\widetilde{N}_{(3,-1)}
= c_0 \widetilde{N}_{(3,-1)}^{c-} + c_0 \xi_0 \widetilde{N}_{(3,-1)}^{c \xi} \,, \quad
\widetilde{N}_{(3,-2)}
= \widetilde{N}_{(3,-2)}^{--} + c_0 \widetilde{N}_{(3,-2)}^{c-}
\end{equation}
with $\widetilde{N}_{(3,-1)}^{c-}$, $\widetilde{N}_{(3,-1)}^{c \xi}$, $\widetilde{N}_{(3,-2)}^{--}$,
and $\widetilde{N}_{(3,-2)}^{c-}$ all annihilated by both $b_0$ and $\eta_0$,
both $\widetilde{N}_{(3,-1)}^{c \xi}$ and $\widetilde{N}_{(3,-2)}^{--}$ impose
the condition $\Phi_{(0,0)}^{c-} = 0$.
So we should be careful if we use
$\widetilde{N}_{(3,-1)}$ and $\widetilde{N}_{(3,-2)}$ as Lagrange multiplier fields.
No such issues arise if we use $N_{(2,-1)}$ with the constraint 
$b_0 \xi_0 N_{(2,-1)} = 0$ as the Lagrange multiplier field.

The Faddeev-Popov term $S_{\rm FP}$ can be obtained from $S_{\rm GF}$
by changing $N_{(2,-1)}$ to the Grassmann-odd antighost field $\Phi_{(2,-1)}$
and by changing $\Phi_{(0,0)}$ to its gauge transformations
$Q \epsilon_{(-1,0)}+\eta_0 \epsilon_{(-1,1)}$
with $\epsilon_{(-1,0)}$ and $\epsilon_{(-1,1)}$ replaced
by the Grassmann-even ghost fields $\Phi_{(-1,0)}$ and $\Phi_{(-1,1)}$, respectively.
We have
\begin{equation}
S_{\rm FP} = \langle \Phi_{(2,-1)} | \,
\Bigl( \, Q | \Phi_{(-1,0)} \rangle + \eta_0 | \Phi_{(-1,1)} \rangle \, \Bigr)
\end{equation}
with the constraint
\begin{equation}
b_0 \xi_0 \Phi_{(2,-1)} = 0 \,,
\end{equation}
which is inherited from $b_0 \xi_0 N_{(2,-1)} = 0$.
After integrating out $N_{(2,-1)}$, the total action we obtain is
\begin{equation}
S_0 + S_1 = -\frac{1}{2} \langle \Phi_{(0,0)} | Q \eta_0 | \Phi_{(0,0)} \rangle
+\langle \Phi_{(2,-1)} | \,
\Bigl( \, Q | \Phi_{(-1,0)} \rangle + \eta_0 | \Phi_{(-1,1)} \rangle \, \Bigr)
\end{equation}
with
\begin{equation}
b_0 \Phi_{(0,0)} = 0 \,, \quad \xi_0 \Phi_{(0,0)} = 0 \,, \quad
b_0 \xi_0 \Phi_{(2,-1)} = 0 \,.
\end{equation}
The action $S_1$ can be written in the following form:
\begin{equation}
S_1 = \langle \, \Phi_{(2,-1)} | \,
\left(
\begin{array}{cc}
Q & \eta_0
\end{array}
\right)
\left(
\begin{array}{c}
\Phi_{(-1,0)} \\
\Phi_{(-1,1)}
\end{array}
\right) \,
\rangle \,.
\end{equation}

This action $S_0+S_1$ is invariant under the following gauge transformations:
\begin{equation}
\begin{split}
\delta_\epsilon \Phi_{(-1,0)} = Q \epsilon_{(-2,0)} +\eta_0 \epsilon_{(-2,1)} \,, \\
\delta_\epsilon \Phi_{(-1,1)} = Q \epsilon_{(-2,1)} +\eta_0 \epsilon_{(-2,2)} \,,
\end{split}
\end{equation}
which can also be written as
\begin{equation}
\delta_\epsilon \left(
\begin{array}{c}
\Phi_{(-1,0)} \\
\Phi_{(-1,1)}
\end{array}
\right) \,
=
\left(
\begin{array}{ccc}
Q & \eta_0 & 0 \\
0 & Q & \eta_0
\end{array}
\right)
\left(
\begin{array}{c}
\epsilon_{(-2,0)} \\
\epsilon_{(-2,1)} \\
\epsilon_{(-2,2)}
\end{array}
\right) \,.
\end{equation}
We can choose $\epsilon_{(-2,1)}$ appropriately
such that the condition
\begin{equation}
\label{vmvmvm}
\xi_0 \Phi_{(-1,0)} = 0
\end{equation}
on $\Phi_{(-1,0)}$ is satisfied. Moreover, we can choose $\epsilon_{(-2,2)}$ appropriately
such that the condition
\begin{equation}
\xi_0 \Phi_{(-1,1)} = 0
\end{equation}
on $\Phi_{(-1,1)}$ is satisfied.
\bigskip
Then $\Phi_{(-1,0)}$ satisfying (\ref{vmvmvm})
can be written as
\begin{equation}
\Phi_{(-1,0)} = \xi_0 \widehat{\Phi}_{(0,-1)} \,,
\end{equation}
where $\widehat{\Phi}_{(0,-1)}$ is in the small Hilbert space.
We can then choose the condition
\begin{equation}
b_0 \widehat{\Phi}_{(0,-1)} = 0  \,,
\end{equation}
to fix the remaining gauge symmetry. 
This gauge-fixing condition can be stated for the original field $\Phi_{(-1,0)}$
as $b_0 \Phi_{(-1,0)} = 0$ when $\xi_0 \Phi_{(-1,0)} = 0$ is imposed.
The condition $b_0 \Phi_{(-1,0)} = 0$
can be satisfied by appropriately choosing $\epsilon_{(-2,0)}$,
and it is compatible with $\xi_0 \Phi_{(-1,0)} = 0$ and $\xi_0 \Phi_{(-1,1)} = 0$
by adjusting $\epsilon_{(-2,1)}$ and $\epsilon_{(-2,2)}$.
To summarize, we can choose
\begin{equation}
b_0 \Phi_{(-1,0)} = 0 \,, \quad \xi_0 \Phi_{(-1,0)} = 0 \,, \quad \xi_0 \Phi_{(-1,1)} = 0 \,,
\label{conditions-on-Phi-1}
\end{equation}
as the gauge-fixing conditions on $\Phi_{(-1,0)}$ and $\Phi_{(-1,1)}$.

Each of $\Phi_{(-1,0)}$ and $\Phi_{(-1,1)}$ can be decomposed into four subsectors
as before so that we have eight subsectors in total.
It is straightforward to see that the conditions~(\ref{conditions-on-Phi-1})
eliminate five of the eight subsectors
and three subsectors remain, which match with the three remaining subsectors
of  $\Phi_{(2,-1)}$
after imposing the constraint $b_0 \xi_0 \Phi_{(2,-1)} = 0$.
We can thus invert the kinetic term $S_1$ to obtain
the propagator,
as we explicitly do in the next section.
It is also straightforward to see that the five conditions can be implemented
by the Lagrange multiplier fields  $N_{(3,-1)}$ and $N_{(3,-2)}$ as
\begin{equation}
S_{\rm GF} = \langle N_{(3,-1)} | \Phi_{(-1,0)} \rangle
+ \langle N_{(3,-2)} | \Phi_{(-1,1)} \rangle \,, 
\end{equation}
with the constraints
\begin{equation}
b_0 \xi_0 N_{(3,-1)} = 0 \,, \quad \xi_0 N_{(3,-2)} = 0 \,.
\end{equation}
This can be verified by decomposing each of $N_{(3,-1)}$ and $N_{(3,-2)}$
into four subsectors.
The corresponding Faddeev-Popov term is then given by
\begin{equation}
S_{\rm FP} = \langle \Phi_{(3,-1)} | \,
\Bigl( Q | \Phi_{(-2,0)} \rangle + \eta_0 | \Phi_{(-2,1) } \rangle \, \Bigr)
+ \langle \Phi_{(3,-2)} | \, 
 \Bigl( Q | \Phi_{(-2,1)} \rangle + \eta_0 | \Phi_{(-2,2) } \rangle \Bigr)
\end{equation}
with
\begin{equation}
b_0 \xi_0 \Phi_{(3,-1)} = 0 \,, \quad \xi_0 \Phi_{(3,-2)} = 0 \,.
\end{equation}
After integrating out $N_{(3,-1)}$ and $N_{(3,-2)}$, the total action we obtain is
\begin{equation}
\begin{split}
S_0 + S_1 + S_2 = & -\frac{1}{2} \langle \Phi_{(0,0)} | Q \eta_0 | \Phi_{(0,0)} \rangle
+\langle \Phi_{(2,-1)} | \,
\Bigl( \, Q | \Phi_{(-1,0)} \rangle + \eta_0 | \Phi_{(-1,1)} \rangle \, \Bigr) \\
& ~+\langle \Phi_{(3,-1)} | \,
\Bigl( Q | \Phi_{(-2,0)} \rangle + \eta_0 | \Phi_{(-2,1) } \rangle \, \Bigr)
+ \langle \Phi_{(3,-2)} | \, 
 \Bigl( Q | \Phi_{(-2,1)} \rangle + \eta_0 | \Phi_{(-2,2) } \rangle \Bigr)
\end{split}
\end{equation}
with
\begin{equation}
\begin{split}
& b_0 \Phi_{(0,0)} = 0 \,, \quad \xi_0 \Phi_{(0,0)} = 0 \,, \qquad \qquad \qquad \qquad \qquad~
b_0 \xi_0 \Phi_{(2,-1)} = 0 \,, \\
& b_0 \Phi_{(-1,0)} = 0 \,, \quad \xi_0 \Phi_{(-1,0)} = 0 \,, \quad \xi_0 \Phi_{(-1,1)} = 0 \,, \qquad
b_0 \xi_0 \Phi_{(3,-1)} = 0 \,, \quad \xi_0 \Phi_{(3,-2)} = 0 \,.
\end{split}
\end{equation}
The action $S_2$ can be written in the following form:
\begin{equation}
S_2 =\bigl \langle \, \left(
\begin{array}{cc}
\Phi_{(3,-1)} & \Phi_{(3,-2)}
\end{array}
\right) |
\left(
\begin{array}{ccc}
Q & \eta_0 & 0 \\
0 & Q & \eta_0
\end{array}
\right)
\left(
\begin{array}{c}
\Phi_{(-2,0)} \\
\Phi_{(-2,1)} \\
\Phi_{(-2,2)}
\end{array}
\right) \, \bigr\rangle \,.    
\end{equation}

The action $S_0+S_1+S_2$
is invariant under the following gauge transformations:
\begin{equation}
\delta_\epsilon \left(
\begin{array}{c}
\Phi_{(-2,0)} \\
\Phi_{(-2,1)} \\
\Phi_{(-2,2)}
\end{array}
\right) \,
=
\left(
\begin{array}{cccc}
Q & \eta_0 & 0 & 0 \\
0 & Q & \eta_0 & 0 \\
0 & 0 & Q & \eta_0
\end{array}
\right)
\left(
\begin{array}{c}
\epsilon_{(-3,0)} \\
\epsilon_{(-3,1)} \\
\epsilon_{(-3,2)} \\
\epsilon_{(-3,3)}
\end{array}
\right) \,.
\end{equation}
It is straightforward to show that we can impose the conditions
\begin{equation}
b_0 \Phi_{(-2,0)} = \xi_0 \Phi_{(-2,0)} = 0 \,, \quad
\xi_0 \Phi_{(-2,1)} = 0 \,, \quad
\xi_0 \Phi_{(-2,2)} = 0
\end{equation}
for gauge fixing.
In this way the gauge-fixing procedure continues, and at the end we obtain
\begin{equation}
S = \sum_{n=0}^\infty \,S_n \,,
\label{cgf action}
\end{equation}
where $S_n$ for $n \ge 1$ is
\begin{equation}
\label{snn>1}
S_n =  \biggl\langle \, 
\left(
\begin{array}{cccc}
\Phi_{(n+1,-1)} & \Phi_{(n+1,-2)} & \cdots & \Phi_{(n+1,-n)}
\end{array}
\right) \bigg| 
\left(
\begin{array}{cccccc}
Q & \eta_0 & 0 & \cdots & 0 & 0 \\
0 & Q & \eta_0 & \cdots & 0 & 0 \\
0 & 0 & Q & \cdots & 0 & 0 \\
\vdots & \vdots & \vdots & \ddots & \vdots & \vdots \\
0 & 0 & 0 & \cdots & Q & \eta_0
\end{array}
\right)
\left(
\begin{array}{c}
\Phi_{(-n,0)} \\
\Phi_{(-n,1)} \\
\vdots \\
\Phi_{(-n,n)}
\end{array}
\right) \, \biggr\rangle 
\end{equation}
with
\begin{equation}
\begin{split}
& b_0 \Phi_{(-n,0)} = \xi_0 \Phi_{(-n,0)} = 0 \,, \qquad
b_0 \xi_0 \Phi_{(n+1,-1)} = 0 \,, \\[1.4ex]  
& \xi_0 \left(
\begin{array}{c}
\Phi_{(-n,1)} \\
\Phi_{(-n,2)} \\
\vdots \\
\Phi_{(-n,n)}
\end{array}
\right) = 0 \,, \qquad \qquad~
\xi_0 \left(
\begin{array}{c}
\Phi_{(n+1,-2)} \\
\Phi_{(n+1,-3)} \\
\vdots \\
\Phi_{(n+1,-n)}
\end{array}
\right) = 0 \,.
\end{split}
\label{Yuji's gauge}
\end{equation}

\subsection{Comparison with Witten's superstring field theory in Siegel gauge}

We have seen that string fields of various ghost and picture numbers appear
in the process of gauge fixing,
and we imposed various conditions
on these string fields.
While those features may look exotic,
we will demonstrate that the gauge-fixed action
of the free superstring field theory in the Berkovits formulation
derived in the preceding subsection
describes the conventional physics
by showing that it reduces to the gauge-fixed action
of the free superstring field theory in the Witten formulation
using Siegel gauge
after eliminating auxiliary fields.

The gauge-invariant action of Witten's superstring field theory is given by
\begin{equation}
\widetilde{S}_0 = -\frac{1}{2} \langle\langle \Psi_{(1,-1)} | Q | \Psi_{(1,-1)} \rangle\rangle \,,
\end{equation}
where $\langle\langle A | B \rangle\rangle$ is the BPZ inner product of $A$ and $B$
in the small Hilbert space, which is related to $\langle A | B \rangle$ in the large Hilbert space
as $\langle\langle A | B \rangle\rangle = (-1)^A \langle A | \xi_0 | B \rangle$
up to an overall sign
depending on a convention.
Here $(-1)^A = 1$ when $A$ is Grassmann even
and $(-1)^A = -1$ when $A$ is Grassmann odd.
Gauge fixing in Siegel gauge is completely parallel
to that in the bosonic string, and the gauge-fixed action is given by
\begin{equation}
\widetilde{S} = \sum_{n=0}^\infty \widetilde{S}_n \,,
\end{equation}
where
\begin{equation}
\widetilde{S}_0 = -\frac{1}{2} \langle\langle \Psi_{(1,-1)} | Q | \Psi_{(1,-1)} \rangle\rangle \,, \qquad
\widetilde{S}_n = -\langle\langle \Psi_{(n+1,-1)} | Q | \Psi_{(-n+1,-1)} \rangle\rangle \quad
\text{for} \quad n \ge 1
\end{equation}
with
\begin{equation}
b_0 \Psi_{(n,-1)} = 0\,, ~~~\forall \,n \,.
\end{equation}
As in the bosonic case, the action $\widetilde{S}$ can also be written compactly as
\begin{equation}
\widetilde{S} = -\frac{1}{2} \langle\langle \Psi | Q | \Psi \rangle\rangle
\quad \text{with} \quad
\Psi = \sum_{n=-\infty}^\infty \Psi_{(n,-1)} \,, \quad b_0 \Psi = 0 \,.
\end{equation}
Since $\Psi$ is annihilated by $b_0$, we need $c_0$ from $Q$
for the inner product to be nonvanishing.
Using $\{ Q, b_0 \} = L_0$, we see that the gauge-fixed action reduces to
\begin{equation}
\widetilde{S} = -\frac{1}{2} \langle\langle \Psi | c_0 L_0 | \Psi \rangle\rangle \,.
\end{equation}
Similarly, $\widetilde{S}_n$ reduces to
\begin{equation}
\widetilde{S}_0 = -\frac{1}{2} \langle\langle \Psi_{(1,-1)} | c_0 L_0 | \Psi_{(1,-1)} \rangle\rangle \,, \qquad
\widetilde{S}_n = -\langle\langle \Psi_{(n+1,-1)} | c_0 L_0 | \Psi_{(-n+1,-1)} \rangle\rangle \quad
\text{for} \quad n \ge 1 \,.
\end{equation}

We will compare this with the gauge-fixed action
derived in the preceding subsection.
Let us start with $S_0$.
Under the gauge-fixing conditions~(\ref{conditions-on-Phi0})
the string field $\Phi_{(0,0)}$ reduces to
\begin{equation}
\Phi_{(0,0)} = \xi_0 \Phi_{(0,0)}^{-\xi} \,.
\end{equation}
Then the action $S_0$ reduces to
\begin{equation}
S_0 = \frac{1}{2} \langle \Phi_{(0,0)}^{-\xi} | \,\xi_0 Q \eta_0 \xi_0 \,| \Phi_{(0,0)}^{-\xi} \rangle
= \frac{1}{2} \langle \Phi_{(0,0)}^{-\xi} | \xi_0 Q | \Phi_{(0,0)}^{-\xi} \rangle
= \frac{1}{2} \langle \Phi_{(0,0)}^{-\xi} | \xi_0 c_0 L_0 | \Phi_{(0,0)}^{-\xi} \rangle \,.
\end{equation}
This coincides with
\begin{equation}
\widetilde{S}_0 = -\frac{1}{2} \langle\langle \Psi_{(1,-1)} | c_0 L_0 | \Psi_{(1,-1)} \rangle\rangle
\end{equation}
in Witten's theory  
under the identification
\begin{equation}
\Psi_{(1,-1)} = \Phi_{(0,0)}^{-\xi} \,.
\end{equation}

Let us next consider $S_1$.
Under the gauge-fixing conditions~(\ref{conditions-on-Phi-1}),
$\Phi_{(-1,0)}$ and $\Phi_{(-1,1)}$ reduce to
\begin{equation}
\Phi_{(-1,0)} = \xi_0 \Phi_{(-1,0)}^{-\xi} \,, \qquad
\Phi_{(-1,1)} = \xi_0 \Phi_{(-1,1)}^{-\xi} + c_0 \xi_0 \Phi_{(-1,1)}^{c \xi} \,.
\end{equation}
Then the action $S_1$ reduces to
\begin{equation}
\begin{split}
S_1 & = \langle \Phi_{(2,-1)} | Q \xi_0 | \Phi_{(-1,0)}^{-\xi} \rangle
+ \langle \Phi_{(2,-1)} | \eta_0 \xi_0 | \Phi_{(-1,1)}^{-\xi} \rangle
+ \langle \Phi_{(2,-1)} | \eta_0 c_0 \xi_0
| \Phi_{(-1,1)}^{c \xi} \rangle \\
& = \langle \Phi_{(2,-1)} | Q \xi_0 | \Phi_{(-1,0)}^{-\xi} \rangle
+ \langle \Phi_{(2,-1)} | \Phi_{(-1,1)}^{-\xi} \rangle
- \langle \Phi_{(2,-1)} | c_0
| \Phi_{(-1,1)}^{c \xi} \rangle \,.
\end{split}
\label{reduced-S_1}
\end{equation}
The antighost field $\Phi_{(2,-1)}$ with the constraint
$b_0 \xi_0 \Phi_{(2,-1)} = 0$ can be decomposed as
\begin{equation}
\Phi_{(2,-1)} = \Phi_{(2,-1)}^{--} + \xi_0 \Phi_{(2,-1)}^{-\xi}
+  c_0 \xi_0 \Phi_{(2,-1)}^{c \xi} \,,
\end{equation}
and the last two terms
on the right-hand side of~(\ref{reduced-S_1})
reduce to
\begin{equation}
\langle \Phi_{(2,-1)} | \Phi_{(-1,1)}^{-\xi} \rangle
= {}-\langle
\Phi_{(2,-1)}^{c \xi} | c_0 \xi_0 | \Phi_{(-1,1)}^{-\xi} \rangle \,, \quad 
\langle \Phi_{(2,-1)} | c_0
| \Phi_{(-1,1)}^{c \xi} \rangle
= {}-\langle
\Phi_{(2,-1)}^{-\xi} | c_0 \xi_0
| \Phi_{(-1,1)}^{c \xi} \rangle \,.
\end{equation}
Since $\Phi_{(-1,1)}^{-\xi}$ and $\Phi_{(-1,1)}^{c \xi}$ 
only appear
in these terms, these fields act as Lagrange multiplier fields imposing
\begin{equation}
\Phi_{(2,-1)}^{c \xi} = 0 \,, \qquad \Phi_{(2,-1)}^{-\xi} = 0 \,.
\end{equation}
After integrating out $\Phi_{(-1,1)}^{-\xi}$ and $\Phi_{(-1,1)}^{c \xi}$, 
the action $S_1$ therefore 
reduces to
\begin{equation}
S_1 = \langle \Phi_{(2,-1)}^{--} | Q \xi_0 | \Phi_{(-1,0)}^{-\xi} \rangle
= \langle \Phi_{(2,-1)}^{--} | c_0 L_0 \xi_0 | \Phi_{(-1,0)}^{-\xi} \rangle \,.
\end{equation}
This coincides with
\begin{equation}
\widetilde{S}_1 = - \langle\langle \Psi_{(2,-1)} | c_0 L_0 | \Psi_{(0,-1)} \rangle\rangle
\end{equation}
in Witten's theory under the identification
\begin{equation}
\Psi_{(2,-1)} = -\Phi_{(2,-1)}^{--} \,, \qquad
\Psi_{(0,-1)} = \Phi_{(-1,0)}^{-\xi} \,.
\end{equation}
We can similarly show that $S_n$ with $n \ge 1$ reduces to
\begin{equation}
S_n = \langle \Phi_{(n+1,-1)}^{--} | Q \xi_0 | \Phi_{(-n,0)}^{-\xi} \rangle
= \langle \Phi_{(n+1,-1)}^{--} | c_0 L_0 \xi_0 | \Phi_{(-n,0)}^{-\xi} \rangle
\end{equation}
and coincides with
\begin{equation}
\widetilde{S}_n = - \langle\langle \Psi_{(n+1,-1)} | c_0 L_0 | \Psi_{(-n+1,-1)} \rangle\rangle
\end{equation}
in Witten's theory under the identification
\begin{equation}
\Psi_{(n+1,-1)} = -\Phi_{(n+1,-1)}^{--} \,, \qquad
\Psi_{(-n+1,-1)} = \Phi_{(-n,0)}^{-\xi}
\qquad \text{for} \quad n \ge 1 \,.
\end{equation}
We have thus shown that the gauge-fixed action derived
in the preceding subsection
coincides with that of Witten's superstring field theory in Siegel gauge
after integrating out auxiliary fields.

While the kinetic term of Witten's superstring field theory
is consistent, 
there are problems in the construction of the cubic interaction term
using the picture-changing operator.
On the other hand, interaction terms can be constructed
without using picture-changing operators in the Berkovits formulation.
We have confirmed that both theories describe the same physics
in the free case, and we expect a regular extension
to the interacting theory in the Berkovits formulation.

\subsection{Various gauge-fixing conditions}
In subsection~\ref{open ssft},
we have seen that the completely gauge-fixed action in the WZW-type open superstring field theory is given by
the sum \eqref{cgf action} of the original action $S_0$
and all the Faddeev-Popov terms 
with the gauge-fixing conditions (\ref{Yuji's gauge}).
As we will 
see later in section \ref{verifying the master eq}, the action (\ref{cgf action}) is precisely the solution to the (classical) master
equation in the BV formalism if we identify antighosts with antifields.\footnote{In the language of the BV formalism 
we have chosen a gauge-fixing fermion such that antifields of minimal-sector fields are identified with antighosts.
In this paper we consider only such gauge-fixing conditions, and thus we will not distinguish antifields and antighosts.}
From this point of view  
$S$ is a universal quantity  
and different gauge-fixed actions can be obtained simply by
imposing different conditions on $\Phi$'s.
In this subsection, we list some gauge-fixing conditions different from (\ref{Yuji's gauge}).
For further generalization and for the validity of the 
gauge-fixing conditions, see~\cite{Shingo}.

Let us first mention
a one-parameter extension of~(\ref{Yuji's gauge}):
\be
\begin{split}
b_0 \Phi_{(-n,0)} &= 0\quad \left( n\geq 0\right),\\
\xi_0 \Phi_{(-n,m)} +\alpha b_0 \Phi_{(-n,m+1)} &= 0\quad \left( 0\leq m\leq n-1\right),\\
\xi_0 \Phi_{(-n,n)} &= 0 \quad \left( n\geq 0\right),\\[2ex]
b_0\xi_0 \Phi_{(n+1,-1)} &= 0 \quad \left( n\geq 1\right),\\
\alpha b_0 \Phi_{(n+1,-m)} + \xi_0 \Phi_{(n+1,-(m+1))} &= 0\quad \left( 1\leq m \leq n-1\right).
\end{split}
\label{one-param bxi}
\ee
The previous condition corresponds to the case in which the parameter $\alpha$ is zero.
Unlike~(\ref{Yuji's gauge}),
the above set of equations includes linear combinations of $\Phi$'s.

Another interesting class of gauge-fixing conditions is obtained when we use  the zero mode $d_0$
of the operator $d = [Q, b\xi]$, instead of $\xi_0$:
\be
\begin{split}
b_0 \Phi_{(-n,0)} &= 0\quad \left( n\geq 0\right),\\
d_0 \Phi_{(-n,m)} +\alpha b_0 \Phi_{(-n,m+1)} &= 0\quad \left( 0\leq m\leq n-1\right),\\
d_0 \Phi_{(-n,n)} &= 0 \quad \left( n\geq 0\right),\\[2ex]
b_0 d_0 \Phi_{(n+1,-1)} &= 0 \quad \left( n\geq 1\right),\\
\alpha b_0 \Phi_{(n+1,-m)} + d_0 \Phi_{(n+1,-(m+1))} &= 0\quad \left( 1\leq m \leq n-1\right).
\end{split}
\label{one-param bd}
\ee
The operator $d$ is identical to the generator $\widetilde{G}^-$ of the twisted $N=2$ superconformal algebra investigated
by Berkovits and Vafa \cite{Berkovits-Vafa}.
Because $d$ is a counterpart of $b$, it seems natural to adopt the symmetric gauge, in which $\alpha = 1$.

In the next section  
we will calculate propagators, mainly considering 
the gauge \eqref{one-param bxi} with $\alpha = 0$, which is identical to \eqref{Yuji's gauge}, 
and the gauge \eqref{one-param bd} with $\alpha = 1$.

\section{Calculation of propagators}
\setcounter{equation}{0}
Let us 
derive propagators under the gauge-fixing conditions proposed in the 
preceding section.
For this purpose, we introduce source terms of the form
\bs
\begin{align}  
S_0^J  
&= \left\langle \Phi_{(0,0)} \,|\, J_{(2,-1)} \right\rangle\,,\\[2ex]
S_n^J    
&=  \sum^{n}_{m=0} \left\langle \Phi_{(-n,m)} \,|\, J_{(n+2,-m-1)}\right\rangle 
+ \sum^{n}_{m=1} \left\langle \Phi_{(n+1,-m)} \,|\, J_{(-n+1,m-1)}\right\rangle
\nonumber \\[2ex]
&= \left\langle
\begin{pmatrix}
\Phi_{(-n,0)} & \cdots & \Phi_{(-n,n)}
\end{pmatrix}
\biggl|
\begin{pmatrix}
J_{(n+2,-1)} \\
\vdots \\
J_{(n+2,-(n+1))}
\end{pmatrix}
\right\rangle
\nonumber \\[1ex]
&\quad +
\left\langle
\begin{pmatrix}
\Phi_{(n+1,-1)} & \cdots & \Phi_{(n+1,-n)}
\end{pmatrix}
\biggl|
\begin{pmatrix}
J_{(-(n-1),0)} \\
\vdots \\
J_{(-(n-1),n-1)}
\end{pmatrix}
\right\rangle
\qquad \left( n\geq 1\right),
\end{align}
\es
and consider the action
\begin{equation}
S_n[J] = S_n + S^J_n 
\quad \left( n\geq 0\right).
\label{S_n J}
\end{equation}
Here $J$'s of positive (non-positive) ghost number are Grassmann-even (Grassmann-odd) sources. 
Each source is coupled with a $\Phi$ of the same Grassmann parity.
Note that $\Phi$'s are 
subject to their gauge-fixing conditions, but sources are free from any constraints.
The actions $S_0$ and $S_n$, for $n\geq 1$, were defined in 
(\ref{s0given}) and (\ref{snn>1}), respectively.
Starting from the action (\ref{S_n J}), we can calculate propagators as follows.
First we solve the equations of motion of the $\Phi$'s derived from $S_n[J]$ in order to find a stationary point.
Then we put the solution back into $S_n[J]$, to obtain a quadratic form of $J$'s, from which propagators can be read off.

\subsection{Propagators for gauge fixing with \texorpdfstring{$b_0$ and $\xi_0$}{b0 and xi0}}
Let us first apply the above-mentioned procedure to the 
gauge (\ref{Yuji's gauge}).
To calculate the propagator of $\Phi_{(0,0)}$ we start from the action
\be
S_0[J] = -\frac{1}{2} \left\langle \Phi_{(0,0)} \,|\, Q \eta_0 \,|\,\Phi_{(0,0)} \right\rangle
+     \left\langle \Phi_{(0,0)} \,|\,J_{(2,-1)} \right\rangle .
\label{S_0 J}
\ee
Gauge-fixing conditions for the 
field $\Phi_{(0,0)}$ are of the form
\be
b_0  \Phi_{(0,0)}  =  \xi_0 \Phi_{(0,0)}  = 0 \,,
\label{p00}
\ee
which leads to
\begin{equation}
\Phi_{(0,0)}  = \{b_0 , c_0 \} \{\xi_0 , \eta_0\} \,\Phi_{(0,0)} = b_0 c_0 \xi_0 \eta_0 \,\Phi_{(0,0)}\,. 
\end{equation}
Thus, the equation of motion derived from $S_0[J]$ is 
\be
\label{eomclgf}
\eta_0 \xi_0 \,  c_0 b_0  \Bigl(  \, Q
\eta_0   
 \,\Phi_{(0,0)} -  J_{(2, -1)} \Bigr) = 0 \,.
\ee
This can be solved easily.
Using the identity
\begin{equation}
Q\eta_0 \frac{\xi_0 b_0}{L_0} = 1 -\frac{b_0 Q}{L_0} -\xi_0\eta_0 + \frac{b_0 Q}{L_0} \xi_0\eta_0\,,
\end{equation}
we find that the solution is given by
\be
\Phi_{(0,0)}  =   {\xi_0 b_0 \over L_0}  \, J_{(2,-1)} \,.
\ee
Note that this solution is consistent with the 
conditions~(\ref{p00}).   
Evaluating the action~(\ref{S_0 J}) for this solution 
determines
the propagator of $\Phi_{(0,0)}$:
\be 
S_0[J] \ = \ {1\over 2} \, \Bigl\langle J_{(2,-1)} \,\Bigl|\, {\xi_0 b_0 \over L_0}\,\Bigr|\, J_{(2,-1)}\Bigr\rangle \,.
\ee

Next, let us consider ghost propagators. The action $S_1 [J]$ 
takes the form
\begin{equation}
\begin{split}  
S_1 [J] \ \, &= \ \ \langle\Phi_{(2, -1)}| \,
\Bigl( Q | \Phi_{(-1,0)}\rangle + \eta_0 | \Phi_{(-1,1) }\rangle\Bigr)\\[1.0ex]
& \ \ + \langle \Phi_{(2,-1)}\,|\,J_{(0,0)}\rangle + \langle \Phi_{(-1,0)}\,|\,J_{(3,-1)}\rangle +\langle \Phi_{(-1,1)}\,|\,J_{(3,-2)} \rangle\,.\
\end{split}
\label{S_1[J]}
\end{equation}
The gauge-fixing conditions at this 
step are
\be
\begin{split}
b_0 \Phi_{(-1,0)} = \xi_0 \Phi_{(-1,0)}  \ &= \ 0 \,, \\
\xi_0 \Phi_{(-1,1)} \ & = \ 0 \,,\\
b_0 \xi_0 \,\Phi_{(2,-1)}  \ & = \ 0 \,.
\end{split}
\label{p2-1}
\ee
Under these conditions, we have
\be
\begin{split}
\Phi_{(-1,0)} &= b_0 c_0 \xi_0 \eta_0 \Phi_{(-1,0)} \,,\\
\Phi_{(-1,1)} &= \xi_0 \eta_0 \Phi_{(-1,1)}\,,\\
\Phi_{(2,-1)} &= \{b_0 , c_0 \} \{\xi_0 , \eta_0\} \,\Phi_{(2,-1)}
= \left( b_0 c_0 + c_0 b_0 \xi_0 \eta_0 \right) \Phi_{(2,-1)}\,.
\end{split}
\ee
Therefore, the equations of motion are  
\be
\begin{split}
\left( c_0 b_0 + b_0 c_0 \eta_0 \xi_0 \right) \left( Q\Phi_{(-1,0)} + \eta_0\Phi_{(-1,1)} + J_{(0,0)} \right) &=0\,,\\
\eta_0 \xi_0 c_0 b_0 \left( Q\Phi_{(2,-1)} + J_{(3,-1)} \right) &=0\,,\\
\eta_0 \xi_0 \left( \eta_0\Phi_{(2,-1)} + J_{(3,-2)} \right) &=0\,.
\end{split}
\ee
Let us
find a solution compatible with the conditions (\ref{p2-1}).
This can be readily achieved by the use of the zero mode decomposition (\ref{decomposition}) of $\Phi$'s.
The solution is given by
\be
\begin{split}
\Phi_{(-1,0)} \ = & \ - {b_0\over L_0} \,\xi_0\eta_0 \, J_{(0,0)} \,, \\
\Phi_{(-1,1)} \ =  & \ \Bigl( -\xi_0   + {b_0\over L_0}  \,\xi_0 \eta_0 X_0 \,\Bigr) J_{(0,0)}\,, \\
\Phi_{(2,-1)} \ = & \  -{b_0\over L_0} \,\eta_0 \xi_0 J_{(3,-1)}+ \Bigl(  -  \xi_0
 + {b_0\over L_0} \, \eta_0 \xi_0 X_0  \Bigr)\, J_{(3,-2)} \,,
\end{split}
\label{sol n=1}
\ee
where $X_0$ is the zero mode of the picture-changing operator 
$X = \{ Q, \xi \}$.
When the equations for $\Phi_{(-1,0)}$ and $\Phi_{(-1,1)}$ hold, the action $S_1[J]$ reduces to 
\begin{equation}
S_1[J] = \langle \Phi_{(2,-1)}\,|\, J_{(0,0)}\rangle  = -\langle J_{(0,0)}\,|\, \Phi_{(2,-1)}\rangle \,.
\label{reduced S1}
\end{equation}
Substituting the solution (\ref{sol n=1}) into (\ref{reduced S1}), we immediately obtain
\be 
S_1[J] \  =  \  \, 
\Bigl\langle J_{(0,0)}\,\Bigl|\,   
{1\over L_0} b_0 \eta_0 \xi_0   \,\Bigr|\, J_{(3,-1)} \Bigr\rangle+ \,\, 
\Bigl\langle J_{(0,0)} 
\,\Bigl|\,
\Bigl( 1 - {1\over L_0} \, b_0 \eta_0 X_0\Bigr) \, \xi_0  \, \Bigr| \, J_{(3,-2)} \Bigr\rangle\,.
\label{result S1}
\ee
On the other hand, when the equation of motion of $\Phi_{(2,-1)}$ holds, the action becomes
\be
S_1[J] =  \langle\Phi_{(-1, 0)}\,|\, J_{(3,-1)}\rangle +  \langle \Phi_{(-1,1)}\,|\, J_{(3,-2)}\rangle\,.
\label{reduced S1'}
\ee
Needless to say, plugging the solution (\ref{sol n=1}) into (\ref{reduced S1'}) gives the same result as in (\ref{result S1}).
The above expression can be rewritten by using a one-by-two propagator matrix:
\be
\label{1af2f-gf2}
S_1[J] \ = \  \Bigl\langle J_{(0,0)} \,\Bigl|\,\begin{pmatrix}
A\ &B
\end{pmatrix}  \begin{pmatrix}  J_{(3,-1)} \\ J_{(3,-2)}  
\end{pmatrix}
\Bigr\rangle
\,,
\ee
with
\be
A \equiv  {1\over L_0} b_0 \eta_0 \xi_0\,, \quad 
B \equiv  \Bigl( 1 - {1\over L_0} \, b_0 \eta_0 X_0\Bigr) \,\xi_0 \,.
\ee 
We emphasize 
that the propagator includes the zero mode of the picture-changing operator.

We can continue the calculation in this manner.
The action  $S_2[J]$  takes the form
\be
\begin{split}
S_2[J]\ =
  & ~~~\, \langle \Phi_{(3, -1)} |\,\Bigl(
  Q | \Phi_{(-2,0)} \rangle + \eta_0 | \Phi_{(-2,1)}\rangle\Bigr)
 + \langle \Phi_{(3,-1)}\,|\,J_{(-1,0)}\rangle\\[0.8ex]
  &  + \langle\Phi_{(3, -2)} |\,\Bigl( 
  Q | \Phi_{(-2,1)}\rangle + \eta_0 | \Phi_{(-2,2)}\rangle\Bigr)
 + \langle \Phi_{(3,-2)}\,|\,J_{(-1,1)}\rangle\\[0.8ex]
 &
 + \langle \Phi_{(-2,0)} \,|\,J_{(4,-1)}\rangle +  \langle \Phi_{(-2,1)} \,|\,J_{(4,-2)}\rangle 
 +\langle \Phi_{(-2,2)} \,
 |\,J_{(4,-3)} \rangle\,,
\end{split}
\ee
and the gauge-fixing conditions are given by
\be
\begin{split}
b_0 \Phi_{(-2,0)} = \xi_0 \Phi_{(-2,0)} &= 0\,,\\
\xi_0 \Phi_{(-2,1)} &=0\,,\\
\xi_0 \Phi_{(-2,2)} &=0\,,\\[1ex]
b_0\xi_0 \Phi_{(3,-1)} &=0\,,\\
\xi_0 \Phi_{(3,-2)} &=0\,.
\end{split}
\ee
When the equations of motion for $\Phi_{(-2,0)}$, $\Phi_{(-2,1)}$, and $\Phi_{(-2,2)}$ are satisfied,
the action $S_2[J]$ reduces~to
\be
S_2[J] = \langle \Phi_{(3,-1)}\,|\,J_{(-1,0)}\rangle + \langle \Phi_{(3,-2)}\,|\,J_{(-1,1)}\rangle
= - \langle J_{(-1,0)}\,|\, \Phi_{(3,-1)}\rangle - \langle J_{(-1,1)}\,|\, \Phi_{(3,-2)}\rangle \,.
\label{reduced S2}
\ee
Substituting into (\ref{reduced S2}) the solution of the equations of motion
\be
\begin{split}
\Phi_{(3,-1)} &= -\left( A \,J_{(4,-1)} +B \,J_{(4,-2)} + (-X_0) B\, J_{(4,-3)} \right),\\
\Phi_{(3,-2)} &= \hspace{31ex} -\xi_0\, J_{(4,-3)}\,,
\end{split}
\ee
we obtain
\be
S_2[J] =
\biggl\langle
\begin{pmatrix}
J_{(-1,0)} & J_{(-1,1)}
\end{pmatrix}
\,\biggl|\,
\begin{pmatrix}
A\  &B  &  (-X_0) B \\ 0 & 0 & \xi_0 
\end{pmatrix} 
\begin{pmatrix}
J_{(4,-1)} \\
J_{(4,-2)} \\
J_{(4,-3)}
\end{pmatrix}
\biggr\rangle
\,.
\ee
At the next step, the propagator matrix 
is given by
\be
 \begin{pmatrix}
A\  &B  &  (-X_0) B & (-X_0)^2 B  \\ 0 & 0 & \xi_0 & (-X_0) \xi_0 \\
0 & 0 & 0 & \xi_0  
\end{pmatrix} ,
\ee 
and
at the $n$-th step
we obtain the $n\times (n+1)$ matrix
\be
\begin{pmatrix}
A& B & (-X_0) B & (-X_0)^2 B & \ldots  &  (-X_0)^{n-1} B \\
0 & 0 & \xi_0 & (-X_0) \xi_0 & \ldots & (-X_0)^{n-2} \xi_0 \\
0 & 0 & 0 & \xi_0&  \ldots & (-X_0)^{n-3} \xi_0 \\
\vdots & \vdots & \vdots & \vdots &  
\ddots & \vdots &\\
0 & 0 & 0 & 0 & \ldots & \xi_0 
\end{pmatrix} .
\label{bxi prop at n}
\ee

The propagators in 
the gauge (\ref{one-param bxi}) with $\alpha\neq 0$ can be calculated in the same manner.
The result, however,  is a little complicated. 
To see this, we calculate the 
first-step ghost propagator.
(Note that since the condition on $\Phi_{(0,0)}$ does not include $\alpha$, 
the propagator of $\Phi_{(0,0)}$ is independent of the parameter.)
We start with the gauge-fixing conditions below:
\be
\begin{split}
b_0 \Phi_{(-1,0)} \ & = \ 0 \,, \\
\xi_0 \Phi_{(-1,0)} + \alpha \, b_0  \Phi_{(-1,1)} \ &= \ 0 \,, \\
\xi_0 \Phi_{(-1,1)} \ & = \ 0 \,,\\
b_0 \xi_0 \,\Phi_{(2,-1)}  \ & = \ 0 \,. \\
\end{split}
\ee
This time, the solution to the equations of motion derived from (\ref{S_1[J]}) is 
\be
\begin{split}
\Phi_{(-1,0)} \ = & \ -\Bigl( \, \frac{\alpha}{1 +\alpha \, L_0} \, b_0 \eta_0 \xi_0
+ {b_0\over L_0} \,\xi_0\eta_0 \, \Bigr) \, J_{(0,0)} \,, \\
\Phi_{(-1,1)} \ =  & \ \Bigl( -\xi_0  + {b_0\over L_0}  \,\xi_0 \eta_0 X_0
+\frac{\alpha}{1 +\alpha \, L_0} \, \xi_0 Q b_0 \eta_0 \xi_0 \,\Bigr) J_{(0,0)} \,, \\
\Phi_{(2,-1)} \ = & \ -\Bigl( \, \frac{\alpha}{1 +\alpha \, L_0} \, b_0 \xi_0 \eta_0 
+ {b_0\over L_0} \,\eta_0\xi_0 \, \Bigr) \, J_{(3,-1)} \\
&\ + \Bigl( -\xi_0  + {b_0\over L_0}  \,\eta_0 \xi_0 X_0
+\frac{\alpha}{1 +\alpha \, L_0} \, \xi_0 \eta_0 b_0 Q \xi_0 \,\Bigr) J_{(3,-2)} \,.
\end{split}
\ee
The action evaluated for the sources is given by
\be
S_1[J] \ = \  \Bigl\langle J_{(0,0)}  \,\Bigl|\,\begin{pmatrix}
A_\alpha\ &B_\alpha
\end{pmatrix}  \begin{pmatrix}  {J_{(3,-1)}} \\ {J_{(3,-2)}}  
\end{pmatrix}\Bigr\rangle\,,
\label{one-param S1J}
\ee
with
\be
A_\alpha \equiv  \frac{\alpha}{1 +\alpha \, L_0} \, b_0 \xi_0 \eta_0 
+ {b_0\over L_0} \,\eta_0\xi_0 \,, \quad 
B_\alpha \equiv  \Bigl( 1 -  {b_0\over L_0}  \,\eta_0  X_0
-\frac{\alpha}{1 +\alpha \, L_0} \, \xi_0 \eta_0 b_0 Q \Bigr) \xi_0 \,.
\ee 
When $\alpha= 0$, the expression (\ref{one-param S1J}) indeed reduces to the form (\ref{1af2f-gf2}).

\subsection{Propagators for gauge fixing with \texorpdfstring{$b_0$ and $d_0$}{b0 and d0}}
Thus far  
we have calculated propagators in the 
gauge \eqref{one-param bxi}, 
focusing on the $\alpha = 0$ case.
These propagators
include the zero mode of the picture-changing operator, which originates from the anticommutation relation
\be
\{ Q, \xi_0\} =X_0\,.
\ee
If instead of $\xi_0$ we use an operator whose anticommutator with $Q$ vanishes, we expect that
propagators are dramatically simplified.
This is indeed the case: the operator $d_0$, 
the zero mode of $d=[ Q,b\xi ]$,
provides us with simpler propagators.
It satisfies the following algebraic relations:
\be
d^2_0 = \{b_0 ,d_0\} = 0\,,\quad 
\{ Q, d_0 \} = 0\,,\quad
\{\eta_0 , d_0 \} =L_0\,.
\ee
In this subsection we investigate propagators in the
gauge (\ref{one-param bd}), 
concentrating on the symmetric case $\alpha=1$.

First we consider $\Phi_{(0,0)}$,
whose gauge-fixing conditions are
\be
b_0  \Phi_{(0,0)}  =  d_0 \Phi_{(0,0)}  = 0 \,.
\label{gfcbd}
\ee
In addition to the source $J_{(2,-1)}$, 
we introduce the Lagrange multipliers $\lambda_{(3,-1)}$ and $\lambda_{(3,-2)}$, and consider the action
$S_0[J] + S_0^\lambda $
with
\begin{equation}
S_0^\lambda =  \bra{ \Phi_{(0,0)} } b_0 \ket{\lambda_{(3,-1)}} + \bra{ \Phi_{(0,0)} } d_0 \ket{\lambda_{(3,-2)}} \,.
\end{equation}
The equation of motion is 
\be
\label{eomclgf2}
-Q\eta_0 \,\Phi_{(0,0)} +  J_{(2, -1)}   +   b_0 \lambda_{(3,-1)}
 +  d_0 \lambda_{(3,-2)}    = 0 \, 
\ee
supplemented by the gauge-fixing conditions \eqref{gfcbd}.
We claim that 
\be
\Phi_{(0,0)}  =   -{b_0\over L_0} \, {d_0 \over L_0}  \, J_{(2,-1)} \,.
\ee
This follows quickly from the identity
\be
Q\eta_0\, {b_0\over L_0} \, {d_0 \over L_0} \ = \   -1  +
 {d_0 \over L_0} \, \eta_0  +  {b_0\over L_0} \,Q + 
{b_0\over L_0} \, {d_0 \over L_0} \, Q\eta_0\,,
\ee
acting on   $J_{(2,-1)}$:  \be
- Q\eta_0\,\Phi_{(0,0)}  \ = \   -J_{(2,-1)}  +
 {d_0 \over L_0} \, \eta_0 \,J_{(2,-1)}  +  {b_0\over L_0} \,Q \,J_{(2,-1)} + 
{b_0\over L_0} \, {d_0 \over L_0} \, Q\eta_0\, J_{(2,-1)} \,. 
\ee
Note that all terms on the right-hand side,
except for the first, simply determine the values of the Lagrange multipliers
in~(\ref{eomclgf2}). Such values are not needed in the evaluation of the action
since the solution satisfies the gauge-fixing conditions.
Evaluating the action for this solution gives
\be 
S_0[J] \ = \ {1\over 2} \, \bigl\langle J_{(2,-1)}\bigl|  \, {d_0\over L_0} \, {b_0 \over L_0}  \, 
\bigr|  J_{(2,-1)}\bigr\rangle\,.
\ee

For the next step we have
the gauge-fixing conditions
\be
\begin{split}
b_0 d_0 \,\Phi_{(2,-1)}  \ & = \ 0 \,, \\
b_0 \Phi_{(-1,0)} \ & = \ 0 \,, \\
d_0 \Phi_{(-1,1)} \ & = \ 0 \,, \\
d_0 \Phi_{(-1,0)} + b_0 \Phi_{(-1,1)}  \ &= \ 0 \,. 
\end{split}
\ee
We implement the first and last gauge conditions with Lagrange multipliers. The relevant action is then 
$S_1[J] + S_1^\lambda$
with
\be 
S_1^\lambda =  - \bra{\lambda_{(4,-2)}}  \Bigl( d_0 \ket{\Phi_{(-1,0)}}  + b_0 \ket{\Phi_{(-1,1)}} \Bigr)
 + \bra{\Phi_{(2,-1)} } b_0 d_0 \ket{\lambda_{(2, -1)}}\,.
\ee
Note that both $J_{(0,0)}$ and $\lambda_{(4,-2)}$ are Grassmann odd.
The gauge-fixed equations of motion are (recall that $d_0$ and $b_0$
are BPZ even, while $\eta_0$ and $Q$ are BPZ odd)
\be
\begin{split}
Q\Phi_{(-1,0)} + \eta_0  \Phi_{(-1,1)}  + J_{(0,0)}   +  b_0 d_0 \lambda_{(2, -1)}
\ & = \ 0 \,,\\
c_0 b_0 \, \Bigl( Q \Phi_{(2,-1)}  + J_{(3,-1)} + d_0 \, \lambda_{(4,-2)} \Bigr)
\ & = \ 0 \,,\\
f_0 d_0 \, \Bigl( \eta_0 \Phi_{(2,-1)}  + J_{(3,-2)} + b_0 \, \lambda_{(4,-2)} \Bigr)
\ & = \ 0 \,,
\end{split}
\ee
where $f_0$ is an operator satisfying $\{d_0,f_0\}=1$.\footnote{
For a concrete expression of $f_0$, see appendix A of \cite{Shingo}.}
In the last equation one may view $J_{(3,-2)} + b_0 \, \lambda_{(4,-2)} $
as a source and solve the equation by writing
\be
\label{phi2-1}
\Phi_{(2,-1)} = - {d_0 \over L_0 }  J_{(3,-2)} \ -\ {d_0\over L_0} b_0 \lambda_{(4,-2)} \ - \ {b_0\over L_0} \eta_0 \, {d_0\over L_0}  J_{(3,-1)}\,,
\ee
where the last term has been included with view of the second equation
and does not disturb the third due to the $\eta_0$ factor it includes.
Substitution into the second equation 
with some simplification yields
\be
c_0 b_0 \, \Bigl( {d_0\over L_0} Q J_{(3,-2)}  + {d_0\over L_0} \eta_0 \,J_{(3,-1)} + 2d_0 \, \lambda_{(4,-2)} \Bigr)\ = \ 0 \,.
\ee
The equation works out if the Lagrange multiplier is given by
\be
\lambda_{(4,-2)} \ = \ -{1\over 2L_0} \bigl(  Q J_{(3,-2)} + \eta_0 J_{(3,-1)} \bigr).
\label{lambda4-2}
\ee
Inserting~\eqref{lambda4-2} 
back in (\ref{phi2-1}),
we now have the solution for $\Phi_{(2,-1)}$.  We find
\be
\begin{split}
\Phi_{(2,-1)} \ = \ -\, {b_0\over L_0}  \, J_{(3,-1)} \ - \  {d_0\over L_0}  \, J_{(3,-2)}
\ + \  {1\over 2}{d_0\over L_0}{b_0\over L_0} QJ_{(3,-2)}
\ + \  {1\over 2}{b_0\over L_0}{d_0\over L_0} \eta_0 J_{(3,-1)}\,.
\end{split}
\ee
A small rearrangement yields 
\be
\begin{split}
\Phi_{(2,-1)} \ = \ -\,{1\over 2} \Bigl(  {b_0\over L_0}  +   {b_0\over L_0} 
\eta_0 {d_0\over L_0}\Bigr)   \, J_{(3,-1)}
-\,{1\over 2} \Bigl(  {d_0\over L_0}  +   {d_0\over L_0} 
Q {b_0\over L_0}\Bigr)   \, J_{(3,-2)}\end{split}\,.
\ee
When the equations for $\Phi_{(-1,0)}$ and $\Phi_{(-1,1)}$ and the 
gauge-fixing conditions hold,  
the action is given by
\be
S_1[J] =  \bra {\Phi_{(2,-1)}} J_{(0,0)} \rangle = - \bra{J_{(0,0)}} \Phi_{(2,-1)}\rangle \,. 
\ee
Its evaluation immediately gives
\be  
S_1[J] \  =  \  \, 
\bigl\langle J_{(0,0)} \bigl|  
\,\frac{1}{2} 
\Bigl(  {b_0\over L_0}  +   {b_0\over L_0} 
\eta_0 {d_0\over L_0}\Bigr)   \, \ket{J_{(3,-1)}}+ \,\, 
\bigl\langle J_{(0,0)} \bigl|
\,\frac{1}{2} 
\Bigl(  {d_0\over L_0}  +   {d_0\over L_0} 
Q {b_0\over L_0}\Bigr)   \, \ket{J_{(3,-2)}}\,.
\ee
This answer can be rewritten by using a one-by-two propagator matrix:
\be
\label{1af2f}
S_1[J] \ = \  
 \Bigl\langle J_{(0,0)}\Bigr| \begin{pmatrix}
\frac{1}{2} \bigl( 
{b_0\over L_0} + {b_0\over L_0} \eta_0 {d_0\over L_0} 
\bigr) 
&~~
\frac{1}{2} \bigl(  
{d_0\over L_0} + {d_0\over L_0}Q {b_0\over L_0}
\bigr) 
\end{pmatrix}  \begin{pmatrix}  J_{(3,-1)} \\ J_{(3,-2)}  
\end{pmatrix}
\Bigr\rangle\,.
\ee
When the equation of motion of $\Phi_{(2,-1)}$ 
and the gauge-fixing conditions hold,  the action reduces~to
\be
S_1[J] =  \bra{\Phi_{(-1, 0)}} J_{(3,-1)}\rangle +  \bra{\Phi_{(-1,1)} } J_{(3,-2)}\rangle\,.
\ee
We can thus read the values of the fields, as bras.  After BPZ conjugation
we obtain 
\be
\begin{split}
\Phi_{(-1,0)} \ = \ &  -{1\over 2} \Bigl(  {b_0\over L_0}  +   {d_0\over L_0} 
\eta_0 {b_0\over L_0}\Bigr)   \, J_{(0,0)} \,,\\[0.5ex]
\Phi_{(-1,1)} \ = \ & -{1\over 2}\Bigl(  {d_0\over L_0}  +   {b_0\over L_0} 
Q {d_0\over L_0}\Bigr)   \, J_{(0,0)} \,.
\end{split}
\ee

\medskip
In the next step we have to deal with three fields and two antifields.  
We have the gauge-fixing conditions
\be
\begin{split}
b_0 \,\Phi_{(3,-1)}  + d_0 \Phi_{(3,-2)}  \ & = \ 0 \,, \\[0.8ex]
b_0 \Phi_{(-2,0)} \ & = \ 0 \,, \\
d_0 \Phi_{(-2,0)} + b_0 \Phi_{(-2,1)}  \ &= \ 0 \,, \\
d_0 \Phi_{(-2,1)} + b_0 \Phi_{(-2,2)}  \ &= \ 0\,, \\
d_0 \Phi_{(-2,2)} \ & = \ 0 \,. 
\end{split}
\ee
The relevant action is 
$S_2[J] + S_2^\lambda$
with
\be
\begin{split}
S_2^\lambda = 
&- \bra{\lambda_{(5,-2)}}  \Bigl( d_0 \ket{\Phi_{(-2,0)}}  + b_0 \ket{\Phi_{(-2,1)}} \Bigr)
- \bra{\lambda_{(5,-3)}}  \Bigl( d_0 \ket{\Phi_{(-2,1)}}  + b_0 \ket{\Phi_{(-2,2)}} \Bigr) \\
&- \bra{\lambda_{(0,0)}} \Bigl( d_0 \ket{\Phi_{(3,-2)}}  + b_0 \ket{\Phi_{(3,-1)}} \Bigr).
\end{split}
\ee
The equations of motion obtained by 
varying the fields, 
\be
\begin{split}
c_0 b_0 \, \Bigl( Q \Phi_{(3,-1)}   + d_0 \, \lambda_{(5,-2)}+ J_{(4,-1)} \Bigr)
\ & = \ 0 \,,\\
Q \Phi_{(3,-2)}  + \eta_0 \Phi_{(3,-1)}
+  b_0 \, \lambda_{(5,-2)}    + d_0 \, \lambda_{(5,-3)}+ J_{(4,-2)}
\ & = \ 0 \,,\\
f_0 d_0 \, \Bigl( \eta_0 \Phi_{(3,-2)}  + b_0 \, \lambda_{(5,-3)} 
 + J_{(4,-3)}\Bigr)
\ & = \ 0 \,,
\end{split}
\ee
are simpler to solve than those obtained by varying the antifields.
By solving the equations for the antifields 
we can determine the Lagrange multipliers:
\be
\begin{split}
\lambda_{(5,-2)} \ &= \ -{1\over 2L_0} \bigl(  Q J_{(4,-2)} + \eta_0 J_{(4,-1)} \bigr)\,, \\
\lambda_{(5,-3)} \ &= \ -{1\over 2L_0} \bigl(  Q J_{(4,-3)} + \eta_0 J_{(4,-2)} \bigr)\,.
\end{split}
\ee
The antifields are then given by 
\be
\begin{split}
\Phi_{(3,-1)} \ &= \ -\,{1\over 2} \Bigl(  {b_0\over L_0}  +   {b_0\over L_0} 
\eta_0 {d_0\over L_0}\Bigr)   \, J_{(4,-1)}
-\,{1\over 2}  {d_0\over L_0}   \, J_{(4,-2)} \,, \\[0.5ex]
\Phi_{(3,-2)} \ &= \ ~~\hskip123pt  - \,{1\over 2}   {b_0\over L_0}    \, J_{(4,-2)}~~
-~\,{1\over 2} \Bigl(  {d_0\over L_0}  +   {d_0\over L_0} 
Q {b_0\over L_0}\Bigr)   \, J_{(4,-3)} \,.
\end{split}\
\ee
Thus the action takes the form 
\be
\label{2af3f}
S_2[J] \ = \  
 \biggl\langle\bigl( J_{(-1,0)} \ J_{(-1,1)} \bigr) \biggr|\begin{pmatrix}
 \frac{1}{2} \bigl( 
{b_0\over L_0} + {b_0\over L_0} \eta_0 {d_0\over L_0} 
\bigr) 
&~~
\frac{1}{2} 
{d_0\over L_0} & 0 \\[0.8ex]
0 & ~~ ~
\frac{1}{2} 
{b_0\over L_0} ~~ &  
\frac{1}{2} \bigl( 
{d_0\over L_0} + {d_0\over L_0}Q {b_0\over L_0}
\bigr) 
~
\end{pmatrix}  \begin{pmatrix}  J_{(4,-1)} \\ J_{(4,-2)} \\  J_{(4, -3)} 
\end{pmatrix} \biggr\rangle . 
\ee
The full pattern is now clear.  The full action $S[J]$ written in terms of propagators and bilinear in sources takes the form
\be
\label{doxioact}    
S[J] \ = \ {1\over 2} \, \bigl\langle J_{2,-1}\bigl|  \, {d_0\over L_0} \, {b_0 \over L_0}  \, 
\bigr|  J_{(2,-1)}\bigr\rangle   +  \sum_{n=0}^\infty  
S_{n+1}[J]\,,
\ee
where $S_{n+1}[J]$  is the term coupling the sources of the
$n+1$ antifields at ghost number $n+2$, to the sources of
the $n+2$ fields at ghost number $-(n+1)$: 
\be
\label{doxioSn}
S_{n+1}[J] \ = \ 
\biggl\langle\Bigl( J_{(-n, 0) }\ J_{(-n, 1)}\   \cdots \ 
J_{(-n, n)}\Bigr) \,\biggl| \,
{\cal P}_{n+1, n+2}  \begin{pmatrix} J_{(n+3, -1)} \\ J_{(n+3, -2)} \\
\vdots \\  J_{(n+3, -(n+2))} \end{pmatrix} \biggr\rangle \,.
\ee
Here the propagator matrix ${\cal P}_{n+1, n+2}$ has $n+1$
rows and $n+2$ columns.   Its general form is the extension
of our results in (\ref{1af2f}) and (\ref{2af3f}):
\be
\label{doxioPk}
\mathcal{P}_{n+1,n+2}
= \frac{1}{2}
\begin{pmatrix}
{b_0\over L_0} + {b_0\over L_0} \eta_0 {d_0\over L_0}  &\frac{d_0}{L_0} &0 &\cdots &0&0&0\\[1ex]
0&\frac{b_0}{L_0} &\frac{d_0}{L_0} &\ddots &\vdots &\vdots &\vdots \\[1ex]
\vdots &0&\frac{b_0}{L_0} &\ddots &0 &\vdots &\vdots \\[1ex]
\vdots &\vdots &0&\ddots &\frac{d_0}{L_0} &0&\vdots \\[1ex]
\vdots &\vdots &\vdots &\ddots &\frac{b_0}{L_0} &\frac{d_0}{L_0} &0 \\[1ex]
0 &0&0&\cdots &0&\frac{b_0}{L_0} & {d_0\over L_0} + {d_0\over L_0}Q {b_0\over L_0}
\end{pmatrix}
\quad \left( n\geq 0\right).
\ee
We can readily obtain the result with a general $\alpha \left( \neq -1\right)$ as well.
The propagator matrices are given by
\begin{subequations}
\begin{align}
&\mathcal{P}_{1,2}
=
\begin{pmatrix}
P_b &P_d
\end{pmatrix}
,\quad
\mathcal{P}_{2,3}
=
\begin{pmatrix}
P_b &\frac{d_0}{(\alpha +1)L_0} &0 \\[1.5ex]
0 &\frac{\alpha b_0}{(\alpha +1)L_0} &P_d
\end{pmatrix}
,\\[3ex]
&
\mathcal{P}_{n+1,n+2}
=
\begin{pmatrix}
P_b &\frac{d_0}{(\alpha +1)L_0} &0 &\cdots &0&0&0\\[1.5ex]
0&\frac{\alpha b_0}{(\alpha +1)L_0} &\frac{d_0}{(\alpha +1)L_0} &\ddots &\vdots &\vdots &\vdots \\[1.5ex]
\vdots &0&\frac{\alpha b_0}{(\alpha +1)L_0} &\ddots &0 &\vdots &\vdots \\[1.5ex]
\vdots &\vdots &0&\ddots &\frac{d_0}{(\alpha +1)L_0} &0&\vdots \\[1.5ex]
\vdots &\vdots &\vdots &\ddots &\frac{\alpha b_0}{(\alpha+1)L_0} &\frac{d_0}{(\alpha +1)L_0} &0 \\[1.5ex]
0 &0&0&\cdots &0&\frac{\alpha b_0}{(\alpha+1)L_0} &P_d
\end{pmatrix}
\quad \left( n\geq 0\right),
\end{align}
\end{subequations}
with
\begin{equation}
P_b = \Bigl( \alpha+\frac{\eta_0 d_0}{L_0} \Bigr)\frac{b_0}{(\alpha +1)L_0}\,,\quad
P_d = \Bigl( 1+\alpha\frac{Q b_0}{L_0} \Bigr)\frac{d_0}{(\alpha +1)L_0}\,.
\end{equation}
Note that the case in which $\alpha = -1$ is exceptional: the propagators diverge, which means that
gauge fixing is not complete. 
See~\cite{Shingo} for details.
When $\alpha = 0$, the above propagators correspond to those obtained from (\ref{bxi prop at n}) 
by the replacement
\be
\xi_0 \longrightarrow \frac{d_0}{L_0}\,,\quad 
X_0 \longrightarrow 0\,.
\ee

\section{Verifying the master equation for the free action}
\label{verifying the master eq}
\setcounter{equation}{0}

Dynamical fields in string field theory are component fields.
The BV formalism is defined in terms of these component fields,
but it is convenient to recast it
in terms of string fields.
In this section we present
the BV formalism of open superstring field theory
in terms of string fields. 
We then show that the free action~(\ref{free-master})
satisfies the master equation.

The classical master equation is given by
\begin{equation}
\label{masterEq}
\{S,S\}=0\,,
\end{equation}
and the antibracket is defined by
\begin{equation}
\label{antibracket987}
\{A,B\}=\sum_k \Big(\frac{\partial_R A}{\partial \phi_k}
   \frac{\partial_L B}{\partial \phi_k^*}-  
   \frac{\partial_R A}{\partial \phi_k^*}
   \frac{\partial_L B}{\partial \phi_k}\Big)\,,
\end{equation}
where $\phi_k$ forms a complete basis of fields and $\phi_k^*$ are the
associated antifields. The Grassmann parity of a field
can be arbitrary but the corresponding antifield
has the opposite parity.  
In our case the fields
are the component fields of $\Phi_-$ and the antifields are
the component fields of $\Phi_+$, 
with $\Phi_\pm$ defined in~\eqref{Phi+-}. 
With a slight abuse of language
we will call $\Phi_-$ the string field and $\Phi_+$ the string
antifield.

Let us expand $\Phi_-$ and $\Phi_+$
in terms of their component fields $f$ and $a$ with indices $g$, $p$, and $r$ as follows:
\begin{subequations}
\label{coefFieldDecomposition}
\begin{align}
&\mbox{String field (even)} &\Phi_- \ =  \sum_{(g,p)\in \Delta_-}
\sum_r f_{g,p}^r\Phi_{g,p}^r\,, ~~~~~~~~~~~~\\
&\mbox{String antifield (odd)} &\Phi_+ \ =  \sum_{(g,p)\in \Delta_+}
\sum_r a_{g,p}^r\Phi_{g,p}^r\,.~~~~~~~~~~~~~
\end{align}
\end{subequations}
We took $f$ and $a$ from the initials of ``fields" and ``antifields."
For each pair $(g,p)$ of the world-sheet ghost number $g$ and picture number $p$,
we chose a complete basis of states $\Phi_{g,p}^r$ labelled by $r$ such that\footnote{
We do not need to consider states with $g=1$,
since they do not appear in the expansion~(\ref{coefFieldDecomposition}).
}
\begin{equation}
\label{ortho}
\langle \, \Phi_{g,p}^r \, | \, \Phi_{g',p'}^{r'} \rangle
\ = \
 \delta_{g+g',2} \,\delta_{p+p',-1}\,\delta_{r,r'} \qquad\qquad g\leq 0\,.
\end{equation}   
Since the Grassmann parity of $\Phi_{g,p}^r$ is $(-1)^g$, we have
\begin{equation}
\label{orthoRev}
\langle \, \Phi_{g,p}^r \, | \, \Phi_{g',p'}^{r'} \rangle
\ = \ (-1)^g\,
 \delta_{g+g',2}\, \delta_{p+p',-1}\,\delta_{r,r'} \qquad\qquad g\geq 2\,,
\end{equation}
which follows from
\begin{equation}
\label{cycleAxiom}
\braket{A}{B}=(-1)^{AB} \braket{B}{A}\,,
\end{equation}
with $(-1)^{gg'} = (-1)^{g(2-g)} = (-1)^{-g^2} = (-1)^g$.
Here and in what follows a string field in the exponent of $(-1)$
represents its Grassmann parity: it is 0 mod 2 for a Grassmann-even string field
and 1 mod 2 for a Grassmann-odd string field.
While the states $\Phi_{g,p}^r$ carry ghost and picture numbers,
the component fields
$f_{g,p}^r$ and $a_{g,p}^r$ do no carry these numbers
and their subscripts $g$ and $p$ refer to
the states that multiply them.
We also introduced
the lattice $\Delta_-$ defined by 
the collection of pairs $(g,p)$ that  appear in $\Phi_-$
and the lattice $\Delta_+$ defined by 
the collection of pairs $(g,p)$ that appear in $\Phi_+$.

As we mentioned in the introduction,
$f_{g,p}^r$ and $a_{2-g,-1-p}^r$ should be paired in the BV formalism:
\be
\hbox{Field-antifield pairing:}~~~f^r_{g,p}  ~ ~\longleftrightarrow 
~~ a^r_{2-g, -1-p} \,.    
\ee
The Grassmann parity of $f_{g,p}^r$ is $(-1)^g$
and that of $a_{g,p}^r$ is $-(-1)^g$.
The Grassmann parity of $a^r_{2-g, -1-p}$ is indeed opposite
to that of $f^r_{g,p}$, which is paired with $a^r_{2-g, -1-p}$.
It then follows that $\Phi_-$ is Grassmann even
and $\Phi_+$ is Grassmann odd.
Note that $a_{g,p}^r$ and
$\Phi_{g,p}^r$ in $\Phi_+$ commute,
while $f_{g,p}^r \, \Phi_{g,p}^r  = (-1)^g  \, \Phi_{g,p}^r \, f_{g,p}^r$ in $\Phi_-$.
The antibracket~(\ref{antibracket987}) is thus defined by
\begin{equation}
\label{antibracket9876}
\{A,B\}=\sum_{(g,p)\in \Delta_-} \sum_r \Big(\frac{\partial_R A}{\partial 
f^r_{g,p}}\,
   \frac{\partial_L B}{\partial a^r_{2-g,-1-p}}\, -  \,  
   \frac{\partial_R A}{\partial a^r_{2-g,-1-p}}\,
   \frac{\partial_L B}{\partial f^r_{g,p}}\Big)\,.
\end{equation}
Our goal is to rewrite this antibracket~(\ref{antibracket9876}) directly
in terms of string fields and string antifields.

In previous sections we used the notation
$\langle \, A \, B \, \rangle$ or $\langle \, A \, | B \, \rangle$
for the BPZ inner product of string fields $A$ and $B$.
When more than two string fields are involved,
it is convenient to introduce the integration symbol as follows:\footnote
{We use the symbol $\star$ to denote the star product in this section.}
\begin{equation}
\int A \star B = \langle \, A \, B \, \rangle = \langle \, A \, | B \, \rangle \,.
\end{equation}
The relation (\ref{cycleAxiom}) is generalized to
\begin{equation}
\label{cycleAxiomInt}
\int A_1 \star A_2 \star\ldots\star A_n =(-1)^{A_1 ( A_2 +\ldots+ A_n)}\int A_2 \star\ldots\star A_n \star A_1 \,.
\end{equation}
The BPZ inner products of states in the basis~(\ref{ortho}) and~(\ref{orthoRev})
are translated into
\begin{subequations}
\label{odlfkd}
\begin{align}
\int\Phi_{g,p}^r\star \Phi_{g',p'}^{r'}\ =&\ \,
 \delta_{g+g',2} \,\delta_{p+p',-1}\,\delta_{r,r'} & g\leq 0\,,\\
\int\Phi_{g,p}^r\star \Phi_{g',p'}^{r'}\ =\ &\,(-1)^g\,
 \delta_{g+g',2} \,\delta_{p+p',-1}\,\delta_{r,r'} & g\geq 2\,.
\end{align}
\end{subequations}

We are interested in evaluating
$\{A,B\}$ where $A$ and $B$ depend
on fields and antifields only through $\Phi_\pm$. 
Let us first consider $\{\Phi_-,\Phi_+\}\,$.
This takes value in a tensor product of two Hilbert spaces of the string field.
We therefore introduce a space number label and write it as $\{\Phi_-^{(1)},\Phi_+^{(2)} \}\,$.
We see that only the first term on the right-hand side of~(\ref{antibracket9876})
contributes and find
\begin{equation}
\label{PhipmAB}
\{\Phi_-^{(1)},\Phi_+^{(2)}\}=\sum_{(g,p)\in\Delta_-} \sum_r
 \frac{\partial_R \Phi_-^{(1)}}{\partial f_{g,p}^r}
   \frac{\partial_L \Phi_+^{(2)}}{\partial a_{2-g,-1-p}^r}\, =\,
   \sum_{(g,p)\in\Delta_-} \sum_r (-1)^g\,
   \Phi_{g,p}^{r(1)}\, \Phi_{2-g,-1-p}^{r(2)}\,,
\end{equation}
where the expansion~(\ref{coefFieldDecomposition}) 
was used and
the sign factor $(-1)^g$ came
from the right derivative that must go through
the state $\Phi_{g,p}^r$ 
to get to the component field $f_{g,p}^r$.  
An important property
of $\{\Phi_-^{(1)},\Phi_+^{(2)}\}$
is that it acts as the  projector $\cP_{\Delta_+}$
to the subspace defined by the lattice $\Delta_+$
in the following sense:
\begin{equation}
\label{prodGivesProj99}
\int_1 X^{(1)} \star_1 
\{\Phi_-^{(1)},\Phi_+^{(2)}\}
 \ = \
 (\cP_{\Delta_+} X)^{(2)}\,,
\end{equation}
where the subscripts attached to the integration symbol
and the star product represent the space number label. 
To see this,
insert (\ref{PhipmAB}) into the left-hand side of (\ref{prodGivesProj99})
\begin{equation}
\label{prodGivesProj}
\int_1 X^{(1)} \star_1 \sum_{(g,p)\in\Delta_-} \sum_r 
(-1)^g \, \Phi_{g,p}^{r(1)}\, \Phi_{2-g,-1-p}^{r(2)} \,=\,
\sum_{(g,p)\in\Delta_-}\hskip-6pt \sum_r (-1)^g   \int_1\bigl( X^{(1)} \star_1 
\, \Phi_{g,p}^{r(1)}\bigr) \Phi_{2-g,-1-p}^{r(2)}
\,,
\end{equation}
and expand
$X^{(1)}$ in a complete basis of ghost and picture numbers,
\be
\label{Xexpand}
X^{(1)} \ = \  \sum_{g',p'=-\infty}^\infty \hskip-3pt \sum_{r'} X^{r'}_{g',p'}  
\Phi^{r' (1)}_{g',p'}\,.
\ee
In~(\ref{prodGivesProj}) this expression is contracted with states
in the subspace defined by $\Delta_-$.
Therefore, in (\ref{Xexpand}) only states
in the subspace defined by $\Delta_+$
give nonvanishing contributions
and the right-hand side of~(\ref{prodGivesProj}) becomes
\be
 \sum_{(g',p')\in\Delta_+}\hskip-3pt \sum_{r'}  X^{r'}_{g',p'}
 \sum_{(g,p)\in\Delta_-}\hskip-6pt \sum_r (-1)^g 
 \int_1 \bigl( \Phi^{r' (1)}_{g',p'} \star_1 \, \Phi_{g,p}^{r (1)}\bigr) 
 \Phi_{2-g,-1-p}^{r(2)} \,.
\ee
Using the second equation in (\ref{odlfkd}), 
we find 
\be
 \sum_{(g',p')\in\Delta_+}\hskip-3pt \sum_{r'}  X^{r'}_{g',p'} 
 \hskip-7pt\sum_{(g,p)\in\Delta_-}\hskip-6pt \sum_r (-1)^g   (-1)^{g'} \delta_{r,r'} 
 \delta_{g+g', 2}\, \delta_{p+p',-1}  \Phi_{2-g,-1-p}^{r(2)} 
 = \sum_{(g',p')\in\Delta_+}\hskip-3pt \sum_{r'}  X^{r'}_{g',p'} 
  \Phi_{g',p'}^{r'(2)} \,.
\ee
The right-hand side is the string field $X^{(1)}$ 
copied into the state space 2, with a projection 
to the subspace defined by $\Delta_+$.
We have thus shown the relation (\ref{prodGivesProj99}). 
Similarly, one can prove
\begin{equation}
\label{projector-Delta-(1)}  
\int_2    \{\Phi_-^{(1)},\Phi_+^{(2)}\}
\star_2 X^{(2)} \ = \
 (\cP_{\Delta_-} X)^{(1)}\,,
\end{equation}
where $\cP_{\Delta_-}$ is the projector
to the subspace defined by $\Delta_-$.
We can also evaluate  $\{\Phi_+^{(1)},\Phi_-^{(2)}\}$ to obtain
\begin{align}
\label{Phi+Phi-}
& \{\Phi_+^{(1)},\Phi_-^{(2)}\} =
  -\sum_{(g,p)\in\Delta_-} \sum_r
   \Phi_{2-g,-1-p}^{r(1)}\, \Phi_{g,p}^{r(2)}\,, \\
\label{Phi+Phi-projectors}   
& \int_1 X^{(1)} \star_1 
\{\Phi_+^{(1)},\Phi_-^{(2)}\}
 \ = \
 -(\cP_{\Delta_-} X)^{(2)}\,, \qquad
 \int_2    \{\Phi_+^{(1)},\Phi_-^{(2)}\}
\star_2 X^{(2)} \ = \
 -(\cP_{\Delta_+} X)^{(1)}\,.
\end{align}

Let us next consider $\{A,\Phi_+ \}$
where $A$ is given by an integral of a product of $\Phi_+$ and $\Phi_-$.
It is useful to define variational derivatives
${\delta_R A\over \delta \Phi_-}$ and ${\delta_R A\over \delta \Phi_+}$ by
\begin{equation}
\label{Avar}
\delta A=\bigint \Bigg(\frac{\delta_R A}{\delta \Phi_-}\star \delta \Phi_- +
              \frac{\delta_R A}{\delta \Phi_+}\star \delta \Phi_+\Bigg)\,.
\end{equation}
We also define ${\delta_L A\over \delta \Phi_-}$ and ${\delta_L A\over \delta \Phi_+}$ by
\begin{equation}
\delta A=\bigint \Bigg( \delta \Phi_- \star \frac{\delta_L A}{\delta \Phi_-}  +
              \delta \Phi_+ \star \frac{\delta_L A}{\delta \Phi_+} \Bigg)\,,
\end{equation}
which are related to ${\delta_R A\over \delta \Phi_-}$ and ${\delta_R A\over \delta \Phi_+}$
as follows:
\begin{subequations}
\label{Lvar-Rvar} 
\begin{align}
\frac{\delta_R A}{\delta \Phi_+} &=
     -(-1)^A \frac{\delta_L A}{\delta \Phi_+}\,,\\  
\frac{\delta_R A}{\delta \Phi_-} &=
      \frac{\delta_L A}{\delta \Phi_-}\,.
\end{align}
\end{subequations}
It is important to note that $\frac{\delta_R A}{\delta \Phi_-}$
and $\frac{\delta_L A}{\delta \Phi_-}$ 
are string fields in the subspace defined by $\Delta_+$
since they are contracted with the variation $\delta \Phi_-$
in the subspace defined by $\Delta_-$.
Similarly, $\frac{\delta_R A}{\delta \Phi_+}$
and $\frac{\delta_L A}{\delta \Phi_+}$
are string fields in the subspace defined by $\Delta_-$
since they are contracted with the variation $\delta \Phi_+$
in the subspace defined by $\Delta_+$.
These can be expressed as follows:
\begin{equation}
\label{string-fields-projected}
\cP_{\Delta_+} \frac{\delta_R A}{\delta \Phi_-} = \frac{\delta_R A}{\delta \Phi_-} \,, \quad
\cP_{\Delta_+} \frac{\delta_L A}{\delta \Phi_-} = \frac{\delta_L A}{\delta \Phi_-} \,, \quad
\cP_{\Delta_-} \frac{\delta_R A}{\delta \Phi_+} = \frac{\delta_R A}{\delta \Phi_+} \,, \quad
\cP_{\Delta_-} \frac{\delta_L A}{\delta \Phi_+} = \frac{\delta_L A}{\delta \Phi_+} \,.
\end{equation}
We can now write 
\begin{equation}
\label{Rderbv}
\frac{\partial_R A}{\partial f_{g,p}^r}
  =\bigint \frac{\delta_R A}{\delta \Phi_-}\star \frac{\partial_R \Phi_-}{\partial f_{g,p}^r}=
		(-1)^g \bigint \frac{\delta_R A}{\delta \Phi_-}\star \Phi_{g,p}^r\,.
\end{equation}
We then obtain
\begin{equation}
\begin{split}  
\{A,\Phi_+^{(2)}\} & =\sum_{(g,p)\in\Delta_-} \sum_r \frac{\partial_R A}{\partial f_{g,p}^r}
   \frac{\partial_L \Phi_+^{(2)}}{\partial a_{2-g,-1-p}^r}=
\bigint_1 \Big(\frac{\delta_R A}{\delta \Phi_-}\Big)^{(1)}\star_1
 \{\Phi_-^{(1)},\Phi_+^{(2)}\} \\
 & = \Big( \cP_{\Delta_+} \frac{\delta_R A}{\delta \Phi_-}\Big)^{(2)} 
 = \Big(\frac{\delta_R A}{\delta \Phi_-}\Big)^{(2)}\,,
\end{split}
\end{equation}
where we used (\ref{PhipmAB}), (\ref{prodGivesProj99}), and (\ref{string-fields-projected}).
Deleting the space number label, we can write the relation as follows:
\begin{equation}
\{A,\Phi_+\} = \frac{\delta_R A}{\delta \Phi_-} \,.
\end{equation}
Similarly, 
one can derive
\begin{equation}
\label{Lderbv}
\begin{split} 
\frac{\partial_R A}{\partial a_{g,p}^r}
  =\bigint \frac{\delta_R A}{\delta \Phi_+}\star \frac{\partial_R \Phi_+}{\partial a_{g,p}^r}=
		\bigint \frac{\delta_R A}{\delta \Phi_+}\star \Phi_{g,p}^r\,, \\
\frac{\partial_L A}{\partial f_{g,p}^r}
  =\bigint \frac{\partial_L \Phi_-}{\partial f_{g,p}^r} \star \frac{\delta_L A}{\delta \Phi_-}=
		\bigint \Phi_{g,p}^r \star \frac{\delta_L A}{\delta \Phi_-} \,, \\
\frac{\partial_L A}{\partial a_{g,p}^r}
  =  
 \bigint 
 \frac{\partial_L \Phi_+}{\partial a_{g,p}^r}   \star \frac{\delta_L A}{\delta \Phi_+}= 
		\bigint  \Phi_{g,p}^r\star \frac{\delta_L A}{\delta \Phi_+}\,,
\end{split}
\end{equation}
as well as the relations
\begin{subequations}
\begin{align}
\{A,\Phi_-\} & =-\frac{\delta_R A}{\delta \Phi_+}\,,\\
\{\Phi_+,A\} & =-\frac{\delta_L A}{\delta \Phi_-}\,,\\
\{\Phi_-,A\} & = \frac{\delta_L A}{\delta \Phi_+}\,.
\end{align}
\end{subequations}
We can see, using (\ref{Lvar-Rvar}), that these relations
are consistent with the familiar property of the 
antibracket:
\begin{equation}
\label{antiBreorder}
\{A,B\}=(-1)^{A+B+AB} \{B,A\} \,.
\end{equation}

Finally, let us consider $\{A, B \}$
where both $A$ and $B$ are integrals of products made of $\Phi_+$ and $\Phi_-$.
In this case, we begin with (\ref{antibracket9876})
and use
(\ref{Rderbv}), (\ref{Lderbv}), (\ref{PhipmAB}), and (\ref{Phi+Phi-}) to show that
\begin{align}
\{A,B\}
=&
  \bigint_{\!\!\!\! 1} \, \bigint_{\!\!\! 2} \, 
  \Bigg(\Big(\frac{\delta_R A}{\delta \Phi_-}\Big)^{(1)}\star_1
	\{\Phi_-^{(1)},\Phi_+^{(2)}\}\star_2\Big(\frac{\delta_L B}{\delta \Phi_+}\Big)^{(2)}
	+
  \Big(\frac{\delta_R A}{\delta \Phi_+}\Big)^{(1)}\star_1
 \{\Phi_+^{(1)},\Phi_-^{(2)}\}\star_2\Big(\frac{\delta_L B}{\delta \Phi_-}\Big)^{(2)}\Bigg)\,.
\end{align}
Using (\ref{prodGivesProj99}) or (\ref{projector-Delta-(1)}),
(\ref{Phi+Phi-projectors}), and (\ref{string-fields-projected}),  we obtain
\begin{equation}
\label{ibracket}
\{A,B\}=
 \bigint\Bigg(\frac{\delta_R A}{\delta \Phi_-}\star \frac{\delta_L B}{\delta \Phi_+}-
              \frac{\delta_R A}{\delta \Phi_+}\star \frac{\delta_L B}{\delta \Phi_-}\Bigg).
\end{equation}
This is the final expression of the antibracket.
The expression (\ref{antibracket9876}) in terms of component fields
has now been written in terms of the string field and the string antifield.

Let us evaluate the antibracket $\{ S, S \}$ for the free action~(\ref{free-master}):
\begin{equation}
\label{free-master2}
S = \int\Big(-\frac{1}{2}\Phi_- \star Q \eta_0 \Phi_-
  + \Phi_+ \star (Q + \eta_0) \Phi_- \Big)\,.  
 \end{equation}
Since 
\begin{equation}
\frac{\delta_R S}{\delta \Phi_-}=
  \cP_{\Delta_+}\Big(-Q\eta_0 \Phi_- +(Q+\eta_0)\Phi_+ \Big)\,, \qquad 
\frac{\delta_L S}{\delta \Phi_+}=\cP_{\Delta_-}\Big((Q+\eta_0)\Phi_- \Big) \,, 
\end{equation}
we find  
\begin{equation}
\frac{1}{2}\{S,S\}=
  \bigint \frac{\delta_R S}{\delta \Phi_-}\star \frac{\delta_L S}{\delta \Phi_+}=
 \bigint \cP_{\Delta_+} \Big(-Q\eta_0 \Phi_- +(Q+\eta_0)\Phi_+ \Big)\star 
   \Big((Q+\eta_0)\Phi_- \Big)  \,.   
\end{equation}
Here we dropped the projector $\cP_{\Delta_-}$
because it is automatically enforced by the other projector $\cP_{\Delta_+}$
through the BPZ contraction.
Since the only string field in $\Phi_-$ such that
$Q\eta_0\Phi_-$ is in the subspace defined by $\Delta_+$ is 
$\Phi_{(0,0)}$ and the action of $Q$ or $\eta_0$ takes string fields
in the subspace defined by $\Delta_+$
to string fields
in the subspace defined by $\Delta_+$,
we have
\begin{equation}
\begin{split}
\frac{1}{2}\{S,S\}\ = \  &
  \bigint \cP_{\Delta_+} \Big(-Q\eta_0 \Phi_- +(Q+\eta_0)\Phi_+ \Big)\star 
  \Big((Q+\eta_0)\Phi_- \Big)  \\ 
  = \  &
  \bigint \Big(-Q\eta_0 \Phi_{(0,0)} +(Q+\eta_0)\Phi_+ \Big)\star 
  \Big((Q+\eta_0)\Phi_- \Big) \,.
  \end{split}
\end{equation}
Using (\ref{etaqprop}),
we conclude that the antibracket $\{ S, S \}$ vanishes
for the free action~(\ref{free-master2}).

\section{Conclusions and outlook}

This paper is the first part of our report on 
the gauge structure and quantization of the
Neveu-Schwarz  superstring field theory~\cite{Berkovits:1995ab}.
A preliminary report of our study of these issues 
was presented by one of the present authors 
(S.T.) at the SFT 2010 conference in Kyoto~\cite{Torii:2011zz}.
In this paper we concentrated on the free theory
and studied a class of gauge-fixing conditions and associated propagators.
We also constructed the free master action and proved that it 
satisfies the classical master equation.  
One could further examine a larger class of
gauge-fixing conditions and associated propagators.
This research direction is described in~\cite{Shingo},
which appears concurrently with this paper.

The next problem is to find the full non-linear master action
for the interacting case.
The result of our study (in which we were joined
by Berkovits) will appear soon in~\cite{PaperII}.
It turned out that it is a difficult problem,
and we have not been able to obtain
a complete form for the master action.
One can think of several approaches
to this problem.
In~\cite{Kroyter:2010rk}, it was shown that a partial
gauge fixing of the cubic democratic theory~\cite{Kroyter:2009rn}
leads to the theory studied here.
If this partial gauge fixing could be extended to the BV
level, it could be used in order to infer the full BV master
action we are after. 
A similar approach, which is presumably simpler,
could be to use another cubic theory  
constructed in such a way as to be
equivalent to the theory we consider here~\cite{SchnablGrassi}.
Being cubic, its BV structure should be simple. If the relation
between the theories could be extended to the BV level, the
master action could be fully written.
A very different approach towards the construction of a master
action is to gauge fix some relatively trivial   
degrees of freedom in a way that leads to a simplified
set of ghosts and antifields.
Such an approach was studied by Berkovits~\cite{Nathan}.
We did not discuss at all in this work the so-called modified
cubic theory~\cite{Preitschopf:fc,Arefeva:1989cp}, but we note
that a quantization of this theory has been proposed very recently
in~\cite{Kohriki:2011zz,Kohriki:2011pp}.\footnote{A discussion of the gauge structure of the modified theory can be found in~\cite{Kroyter:2009zi}.}

The complete open superstring field theory
includes both Neveu-Schwarz and Ramond sectors.
In the current discussion we ignored the Ramond sector.
Its inclusion  is certainly an important
goal. It might be possible to generalize the current
construction to the Ramond sector using ideas
from~\cite{Berkovits:2001im,Michishita:2004by,Kroyter:2009rn,Kroyter:2010rk}.
We leave the incorporation of the Ramond sector for future work.

\section*{Acknowledgments}

\noindent

We would like to thank Stefan Fredenhagen, Leonardo Rastelli, Ashoke Sen, and Warren Siegel
for useful discussions.
We are especially grateful to Nathan Berkovits for discussions and
collaboration on these and related matters.
The work of M.K. was supported by an Outgoing
International Marie Curie Fellowship of the European Community. The views presented
in this work are those of the authors and do not necessarily reflect those of the European
Community.
The work of Y.O. was supported in part by Grant-in-Aid for Young Scientists~(B) No.~21740161
from the Ministry of Education, Culture, Sports, Science and Technology (MEXT) of Japan 
and by Grant-in-Aid for Scientific Research~(B) No.~20340048 from the Japan Society for the Promotion
of Science (JSPS). 
The work of M.S. was supported by the EURYI grant GACR EYI/07/E010 from EUROHORC and ESF.
The work of S.T. was supported in part by Research Fellowships of JSPS for Young Scientists.
The work of Y.O., M.S. and S.T. was also supported in part
by the M\v{S}MT contract No. LH11106 and 
by JSPS and the 
Academy of Sciences of the Czech Republic (ASCR)
under the Research Cooperative Program between Japan and the Czech Republic.
The work of B.Z. was supported in part by the U.S. Department of Energy (DOE)
under the cooperative research agreement DE-FG02-05ER41360.

\end{document}